\documentclass[sigconf]{acmart}
\AtBeginDocument{%
  }

\usepackage{multirow}
\usepackage{adjustbox}
\usepackage{booktabs}
\usepackage{placeins}
\usepackage{algorithm}
\usepackage{algpseudocode}
\usepackage{subcaption}
\setcopyright{acmlicensed}
\acmYear{2018}
\acmDOI{XXXXXXX.XXXXXXX}
\begin{document}


\title{AUDETER: A Large-scale Dataset for Deepfake Audio Detection in Open Worlds}

\author{Qizhou Wang}
\email{mike.wang@unimelb.edu.au}
\affiliation{%
  \institution{The University of Melbourne}
  \city{Parkville}
  \country{Australia}
}

\author{Hanxun Huang}
\email{hanxun@unimelb.edu.au}
\affiliation{%
  \institution{The University of Melbourne}
  \city{Parkville}
  \country{Australia}
}

\author{Guansong Pang}
\email{gspang@smu.edu.sg}
\affiliation{%
  \institution{Singapore Management University}
  \country{Singapore}
}

\author{Sarah Erfani}
\email{sarah.erfani@unimelb.edu.au}
\affiliation{%
  \institution{The University of Melbourne}
  \city{Parkville}
  \country{Australia}
}

\author{Christopher Leckie}
\email{caleckie@unimelb.edu.au}
\affiliation{%
  \institution{The University of Melbourne}
  \city{Parkville}
  \country{Australia}
}

\renewcommand{\shortauthors}{Wang et al.}

\begin{abstract}
Speech synthesis systems can now produce highly realistic vocalisations that pose significant authenticity challenges. Despite substantial progress in deepfake detection models, their real-world effectiveness is often undermined by evolving distribution shifts between training and test data, driven by the complexity of human speech and the rapid evolution of synthesis systems. Existing datasets suffer from limited real speech diversity, insufficient coverage of recent synthesis systems, and heterogeneous mixtures of deepfake sources, which hinder systematic evaluation and open-world model training.
To address these issues, we introduce AUDETER (AUdio DEepfake TEst Range), a large-scale and highly diverse deepfake audio dataset comprising over 4,500 hours of synthetic audio generated by 11 recent TTS models and 10 vocoders, totalling 3 million clips. We further observe that most existing detectors default to binary supervised training, which can induce negative transfer across synthesis sources when the training data contains highly diverse deepfake patterns, impacting overall generalisation. As a complementary contribution, we propose an effective curriculum-learning-based approach to mitigate this effect. Extensive experiments show that existing detection models struggle to generalise to novel deepfakes and human speech in AUDETER, whereas XLR-based detectors trained on AUDETER achieve strong cross-domain performance across multiple benchmarks, achieving an EER of 1.87\% on In-the-Wild. AUDETER is available on \href{https://github.com/mike-qz-wang/AUDETER}{\textcolor{blue}{GitHub}}.
\end{abstract}

\received{8 February 2026}

\maketitle

\section{Introduction}
Deepfake audio detection (DFAD) is the task of identifying audio generated by speech synthesis models, such as Text-To-Speech (TTS) systems and vocoders. There has been a long history of developing DFAD models due to the significance of its real-world applications such as authentication in forensics, misinformation detection on social media, and voice biometric security systems. 

\begin{figure}
    \centering
    \includegraphics[width=0.8\columnwidth]{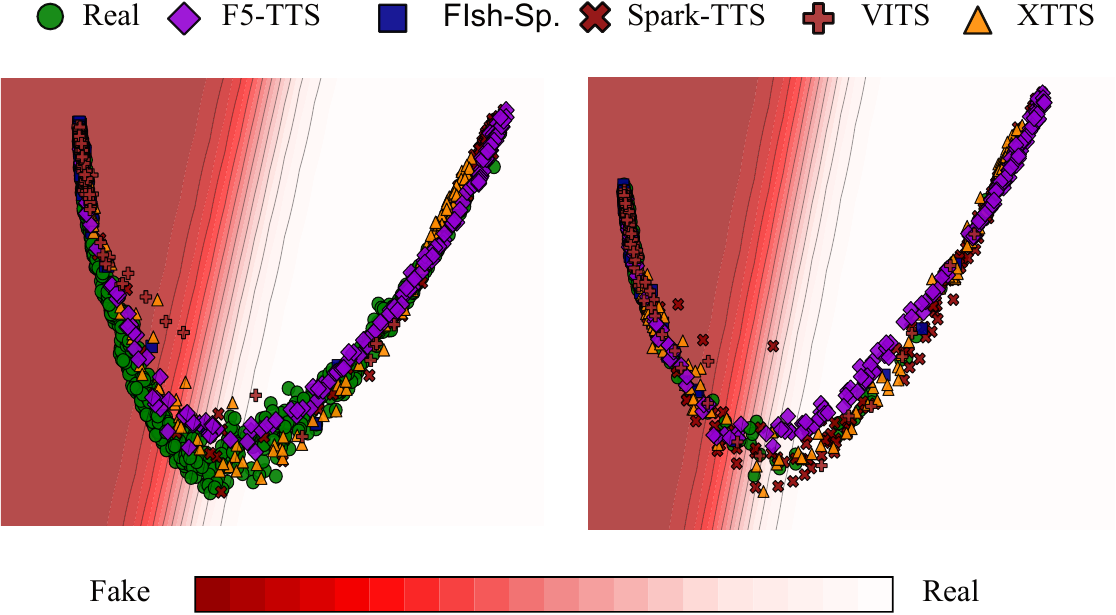}
    \caption{UMAP visualisation of last-layer audio representations from an XLR+R+A \cite{wra} model trained on ASVSpoof and evaluated on AUDETER. Colours indicate normalised real-class likelihood. The strong overlap between real samples and five deepfake types, including many real samples in high-spoof regions, showing limited generalisation and both false positives and false negatives.}
    \label{fig:intuition}
\end{figure}
State-of-the-art (SOTA)  DFAD models have shown strong performance against current benchmark datasets, such as ASVSpoof \cite{wang2020asvspoof,yamagishi2021asvspoof,wang2024asvspoof} and In-the-Wild (ITW) \cite{itw}. However, their performance in real-world scenarios remains unreliable, particularly when detecting deepfake audio generated by novel speech synthesis systems not seen during training, as well as human voices with different acoustic characteristics and artefacts. This is because most DFAD models are trained under a closed-world binary classification formulation, which fits only the limited acoustic and synthetic patterns present in the training data and consequently degrades open-world generalisation, as illustrated in Fig.~\ref{fig:intuition}. We refer to {\it open-world detection} as the setting where both real and synthetic audio samples at inference time may originate from previously unseen distributions or synthesis systems not explicitly included in training.

Current datasets are becoming increasingly limited in covering diverse and recent deepfake speech patterns. As a result, they fail to adequately challenge modern DFAD models for open-world detection.
To improve open-world DFAD evaluation and model training, we introduce \textbf{AUDETER} (AUdio DEepfake TEst Range), a large-scale deepfake audio detection dataset that consists of audio samples systematically created using a wide variety of speech synthesis models and multiple sources of human voices. Table \ref{tab:compare} compares AUDETER with the existing datasets. AUDETER contains 4,682 hours of deepfake audio generated using 21 speech synthesis systems, including 10 recent TTS models, corresponding to 4 human voice corpora, making it substantially larger and more diverse than previous datasets. In particular, each real audio sample is paired with matched synthetic versions from all 21 synthesis systems, enabling balanced evaluation and controlled analysis of domain shifts in open-world detection. Moreover, the quality of all generated audio is rigorously assessed for intelligibility and naturalness (see Fig.~\ref{fig:gen_pipeline} for AUDETER’s audio generation pipeline).

\begin{table}[t]
\caption{A comparison between our proposed AUDETER dataset and existing datasets.}
\begin{adjustbox}{width=\columnwidth, center}
\begin{tabular}{ccccccc} \toprule
Dataset & \begin{tabular}[c]{@{}c@{}}\# Audio\\Clips\end{tabular} & \begin{tabular}[c]{@{}c@{}}Total\\Hours\end{tabular} & \begin{tabular}[c]{@{}c@{}}Diverse\\Real\end{tabular} & \begin{tabular}[c]{@{}c@{}}\# Synthetic\\Models\end{tabular} & \begin{tabular}[c]{@{}c@{}}\# TTS\\Models\end{tabular} & \begin{tabular}[c]{@{}c@{}}\# Matching\\Script\end{tabular} \\
\hline
ASV 2019 \cite{wang2020asvspoof} &312K  &60-70 &N &19 &N &N \\
ASV 2021 DF \cite{yamagishi2021asvspoof} &612K &100-120 &N &13 &N  &N \\
In-the-Wild \cite{itw} & 31.8K  &17.2 &N  &-  & N &N \\
WaveFake  \cite{frank2021wavefake} &137K &175 & N & 6 &N &Y\\
LibriSeVoc \cite{cvprw} & - &126.41 &N &6 &N &Y \\ 
DFADD \cite{du2024dfadd} &1K  &221 &N &5 &Y &N  \\
SONAR \cite{sonar} &2K  &8 &N &8 &Y &N  \\\cmidrule(lr){1-7} 
AUDETER (ours) &3M &4,681.9 &4 &21 &11 &Y \\
\bottomrule 
\end{tabular}
\end{adjustbox}
\label{tab:compare}
\end{table}
AUDETER not only serves as a challenging and systematic evaluation benchmark, but a valuable resource that offers diverse coverage at scale for training DFAD models. For evaluation, our extensive experiments on AUDETER reveal significant performance degradation of models trained on existing datasets that include recent, novel systems, underscoring the limitations of current benchmarks and training data. AUDETER’s consistent and uniform structure enables systematic cross-system analysis, providing a flexible foundation for future investigations into DFAD generalisation. We present representative findings in Section~\ref{sec:training}.

For training, detection models trained on AUDETER consistently outperform their pretrained counterparts under the same training setup, as well as leading baseline detectors, in both in-domain and cross-dataset evaluations. This highlights AUDETER’s comprehensive and representative data coverage and its effectiveness as a data-centric approach to improving detection performance. In addition, AUDETER-trained models exhibit increased robustness to real-world challenges such as background noise. Specifically, using a SOTA DFAD model architecture, XLR-SLS~\cite{mm}, we retrain the detector on AUDETER and achieve an Equal Error Rate (EER) of 1.87\% on the ITW dataset, demonstrating that AUDETER provides sufficient training coverage for data-demanding backbone models.

As a complementary contribution, we identify a key limitation of the default closed-world supervised training approach commonly used by existing detection models, which treats all deepfake audio as a single homogeneous class without explicitly modelling differences across speech synthesis systems. Enabled by AUDETER’s diverse synthesis patterns, we find that directly training under this formulation on highly diverse speech models can induce negative transfer, degrading cross-domain performance. To further study this effect, we extend cross-domain evaluation to two additional recent datasets, revealing that systems with strong synthesis fingerprints can disproportionately harm overall cross-domain generalisation. To alleviate this issue, we cast learning from diverse deepfake data as a curriculum learning problem implemented via a two stage pipeline. In stage one, we identify systems with strong synthesis fingerprints that often dominate the training signal using a linear probing test and train a base detector on the remaining data to learn universal, system invariant representations. In stage two, the model is adapted to all speech systems, with the stage one backbone used as a teacher to regularise adaptation and preserve general cues that cannot be enforced through data labels. This leads to balanced and substantially improved cross-domain performance across multiple benchmarks. Our main contributions are as follows:
\begin{itemize}
\item We introduce AUDETER, a large-scale deepfake audio detection dataset comprising 4,682 hours of synthetic audio generated by us using 21 recent speech synthesis systems across four human voice corpora.  AUDETER provides substantially greater scale and diversity than existing datasets, supporting improved evaluation and model training for open-world detection.
\item Through extensive evaluation, AUDETER is shown to effectively challenge existing models, exposing their limited suitability for open-world detection. As a training resource, AUDETER enables significant improvements over prior SOTA performance.
\item We identify a key limitation of closed-world training where dominant synthesis fingerprints hinder generalisation, and address this with a curriculum learning based method that significantly improves cross-domain performance.
\end{itemize}

\section{Related Work and Preliminaries}

\subsection{Related Work}
\subsubsection{Deepfake Audio Detection.} DFAD models have been extensively studied \cite{wang2020asvspoof,yamagishi2021asvspoof,frank2021wavefake,yi2022add,itw,cvprw,wang2024asvspoof,zhang2025i,m2, hiercon}. Earlier DFAD models explore various types of features, such as short-term spectral \cite{xiao2015spoofing, tian2016spoofing}, long-term spectral \cite{alegre2013new, alegre2013one, sahidullah2015comparison}, prosodic \cite{pal2018synthetic, xue2022audio, wang2023detection}, and deep features \cite{yu2017dnn, sailor2017unsupervised, cheuk2020nnaudio, fu2022fastaudio, zeghidour2018learning, ravanelli2018speaker}. A wide range of deep neural networks (DNN)  architectures \cite{wu2020light,liu2023leveraging,rawnet2, jung2022pushing,ge2021raw} have also been studied for enhanced feature extraction.
More recent works employ pretrained audio models such as Wav2Vec~2.0 \cite{baevski2020wav2vec} as backbones, leveraging their greater capacity and extensive audio-domain pretraining for improved representation learning. These backbones \cite{babu2021xls,conneau2020unsupervised} are commonly paired with DNNs as scoring models and represent the current state-of-the-art \cite{yang2024robust,wra,mm}. However, these models are trained as closed-set classifiers and often fail to generalise to novel samples at inference time.

\subsubsection{Speech Synthesis Models}
\textbf{End-to-end TTS Systems.} TTS systems convert text input to audio waveforms through neural networks. Earlier approaches like Tacotron \cite{wang2017tacotron}, WaveNet \cite{oord2016wavenet}, and FastSpeech \cite{ren2019fastspeech} follow a two-stage workflow, in which text inputs are first converted to acoustic features, which are then processed by vocoders to generate waveforms. More sophisticated generative models are used to improve this process such as flow-based generation (Glow-TTS) \cite{glow} and end-to-end variational approaches (VITS) \cite{vits}. More recent models such as YourTTS \cite{yourtts} and OpenVoice \cite{ov} enable voice cloning with basic emotion and style control. However, these methods rely on phonetic alignments for generation and suffer from limited emotional realism and poor prosodic modelling capabilities such as intonation. More recent TTS systems \cite{cosyvoice, sparktts, liao2024fish, bark, vits} leverage large language models and large-scale pretraining, significantly improving speech quality and realism.
\smallskip

\noindent \textbf{Vocoders.} Vocoders convert intermediate acoustic features (such as mel-spectrograms) into  audio waveforms. Traditional methods \cite{morise2016world, degottex2011glottal}   used signal processing techniques, but neural vocoders have substantially improved audio quality. Early neural approaches \cite{oord2016wavenet, mehri2016samplernn} use autoregressive generation but suffer from slow inference. Subsequent developments like  \cite{prenger2019waveglow} and Parallel WaveGAN \cite{pwg} introduced parallel generation for faster synthesis. GAN-based vocoders such as MelGAN~\cite{melgan}, HiFi-GAN \cite{highgan}, and UnivNet \cite{univnet} further improved efficiency and quality through adversarial training. More recent vocoders like BigVGAN \cite{bigvgan} and Vocos \cite{vocos} have incorporated advanced architectures and training techniques to achieve state-of-the-art audio fidelity. Modern vocoders can generate audio that is nearly indistinguishable from real human speech.

\subsubsection{Deepfake Audio Datasets}
ASVSpoof series datasets~\cite{wang2020asvspoof, yamagishi2021asvspoof} are widely used for deepfake audio detection. ASVSpoof 2019 focuses on TTS and voice conversion attacks, while ASVSpoof 2021 expanded to include more diverse spoofing methods.In-the-Wild \cite{itw} is a common choice for evaluating cross-domain detection, but its small scale and unspecified generation methods make it unsuitable for training and fine-grained evaluation. WaveFake \cite{frank2021wavefake} and LibriSeVoc focus mainly on vocoder-based synthesis and remain limited in scale. DFADD \cite{du2024dfadd} and SONAR \cite{sonar} include multiple TTS systems but are primarily evaluation-focused and not designed for large-scale training.

\subsection{Problem Statement}
\label{sec:ps}
Deepfake audio detection aims to learn a scoring function \( S(\cdot) \) that assigns authenticity scores to audio samples, such that \( S(x_{\text{fake}}) \ll S(x_{\text{real}}) \). A decision threshold \( \tau \) is applied to classify samples as real if \( S(x) > \tau \) and fake otherwise. Detection performance is commonly evaluated using the Equal Error Rate (EER), defined as the error rate at the decision threshold where the false acceptance rate (FAR) equals the false rejection rate (FRR). Lower EER indicates better detection performance. \textit{Most existing methods implement $S$ using deep neural networks learned through supervised training, while open-world deepfake detection requires generalisation to unseen synthesis systems and diverse human speech styles.}

\section{AUDETER Dataset}

\begin{figure}[t]
    \centering
    \caption{UMAP visualisation of real and synthetic audio samples. AUDETER captures more diverse real speech patterns and a broader range of synthetic variations compared to existing datasets.}
    \begin{subfigure}[t]{0.48\linewidth}
        \centering
        \includegraphics[width=\linewidth]{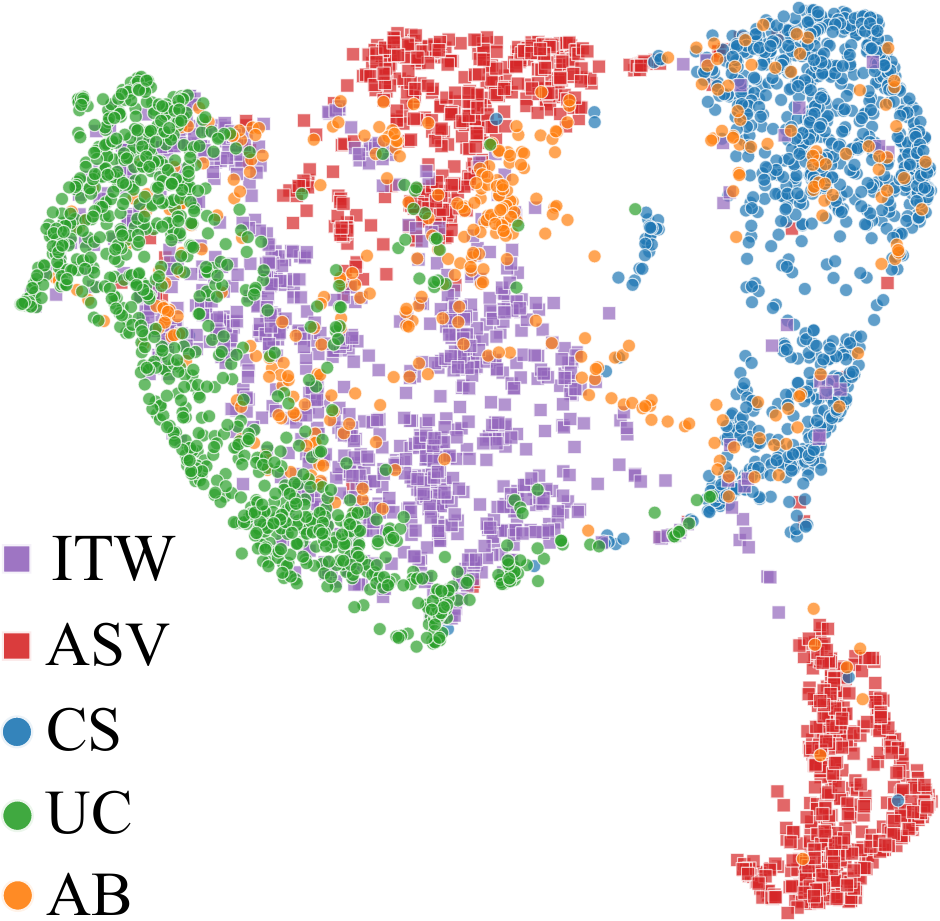}
        \caption{Real speech samples from CS, UC AB subsets in AUDETER, compared with ASVSpoof 2021 DF and In-the-Wild (ITW).}
        \label{fig:umap_real}
    \end{subfigure}
    \hfill
    \begin{subfigure}[t]{0.48\linewidth}
        \centering
        \includegraphics[width=\linewidth]{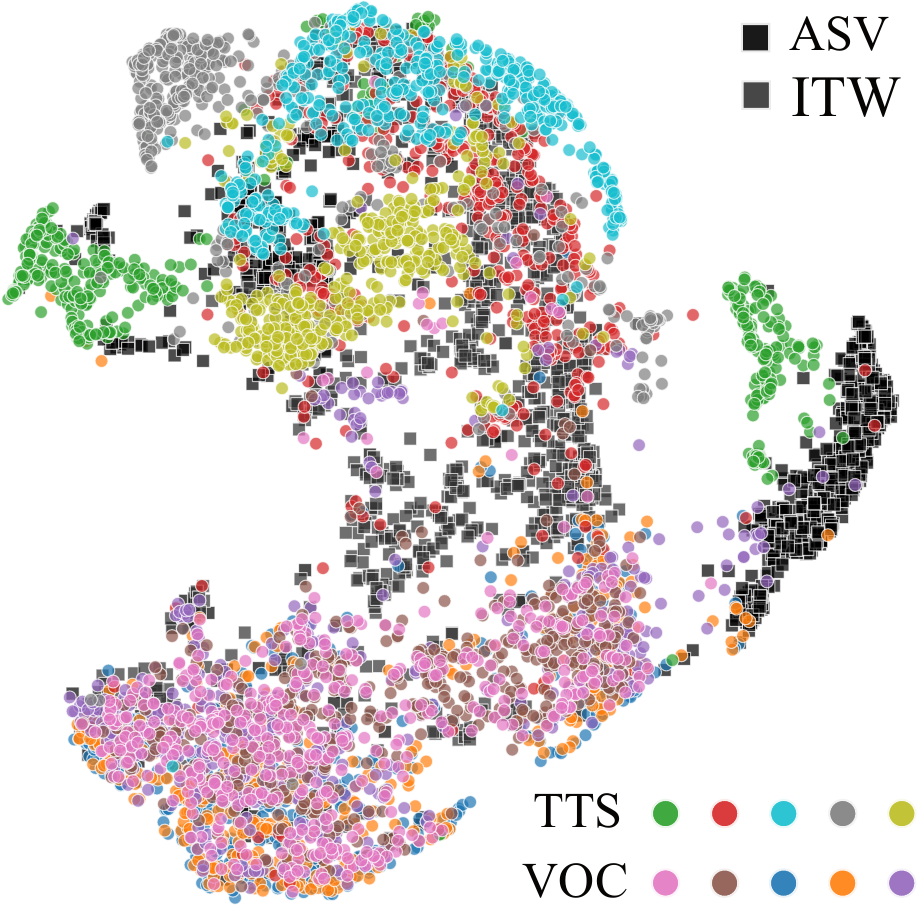}
        \caption{Deepfake audio samples from selected speech systems in AUDETER, compared with ASVSpoof 2021 DF and ITW.}
        \label{fig:umap_fake}
    \end{subfigure}
    \label{fig:umap}
\end{figure}

\subsection{Design Motivations}
AUDETER is designed to address key limitations in existing deepfake audio benchmarks, both for evaluating open-world detection and for supporting the training of generalised detectors at scale.

\subsubsection{Enhanced Open-world Evaluation} 
Existing datasets provide limited coverage of recent speech synthesis methods and lack sufficient diversity in real speech. As shown in Section~\ref{sec:testdata}, this leads to significant performance degradation under novel synthesis systems and human voice domain shifts. To address this, AUDETER includes synthetic audio from 21 recent TTS and vocoder models, paired with real speech from four large-scale corpora covering diverse speaking styles and conditions. Consistent with our open-world definition, AUDETER is structured to allow controlled evaluation under both unseen synthesis systems and unseen real-speech distributions.

\subsubsection{Towards Training Generalist Detection Models.} Existing models struggle under domain shifts due to limited training data. A data-centric approach to generalised detection requires large-scale training data capturing diverse real and deepfake patterns, which is fundamental for effective model learning and is provided by AUDETER. Beyond scale and diversity, AUDETER’s balanced and structured design enables systematic study of training strategies for open-world generalisation under controlled distribution shifts.

\subsection{Dataset Overview}
\begin{table}[t]
\caption{The organisation of the AUDETER dataset.}
\begin{adjustbox}{width=0.9\columnwidth}
\begin{tabular}{cccccc} \toprule
Collection               & Subset                         & Partition & Patterns & \# Audio / Patt. & Total Hrs \\ \midrule
\multirow{7}{*}{TTS}     & Celebrity (\textbf{CE})                    &Test & 15       & 19,784            & 311.5     \\
                         & \multirow{2}{*}{CrowdSource (\textbf{CS})}  & Train      & 15       & 16,372            & 275.0     \\
                         &                                & Test      & 15       & 16,372            & 265.3     \\
                         & \multirow{2}{*}{US Congress (\textbf{UC})} & Train       & 15       & 18,622            & 493.5     \\
                         &                                & Test      & 15       & 34,898            & 909.4     \\
                         & \multirow{2}{*}{Audiobook (\textbf{AB})}           & Train       & 15       & 3,807             & 212.7     \\
                         &                                & Test      & 15       & 3,769             & 209.1     \\ \midrule
\multirow{7}{*}{Vocoder} & Celebrity (\textbf{CE})                 & Test & 10       & 19,784            & 207.6     \\
                         & \multirow{2}{*}{CrowdSource (\textbf{CS})}  & Train       & 10       & 16,372            & 266.7     \\
                         &                                & Test      & 10       & 16,372            & 264.8     \\
                         & \multirow{2}{*}{US Congress (\textbf{UC})} & Train       & 10       & 18,622            & 331.7     \\
                         &                                & Test      & 10       & 34,898            & 598.1     \\
                         & \multirow{2}{*}{Audiobook (\textbf{AB})}           & Train       & 10       & 3,807             & 156.7     \\
                         &                                & Test      & 10       & 3,769             & 154.9    \\ \bottomrule
\end{tabular}
\end{adjustbox}
\label{tab:dset_org}
\end{table}

\begin{table}[t]
\caption{A list of the TTS systems and vocoders employed to produce the AUDETER dataset.}
\begin{adjustbox}{width=\columnwidth,center}
\begin{tabular}{p{1.6cm}p{9cm}}
\toprule
TTS Syst. & CosyVoice (2025) \cite{cosyvoice}, Zonos (2025) \cite{zonos}, SparkTTS (2025) \cite{sparktts}, F5-TTS (2025) \cite{chen2024f5}, Fish-Speech (2024) \cite{liao2024fish}, OpenVoice V2 (2023) \cite{ov}, ChatTTS (2024) \cite{chattts}, XTTS v2 (2024) \cite{xtts}, Bark (2023) \cite{bark}, YourTTS (2022) \cite{yourtts}, VITS (2021) \cite{vits} \\ 
\midrule
Vocoders    & BigVGan (2022) \cite{bigvgan}, BigVSan (2024) \cite{bigvsan}, Vocos \cite{vocos} (2023), UnivNet (2021) \cite{univnet}, HiFi-GAN (2020) \cite{highgan}:, MelGAN (2019) \cite{melgan} , Full-band MelGAN \cite{parallelwavegan_lib}, Multi-band MelGAN \cite{mbmel}, Parallel WaveGAN \cite{pwg}, Style MelGAN \cite{stylemel}.\\
\bottomrule
\end{tabular}
\end{adjustbox}\label{tab:syn_models}
\end{table}

Table~\ref{tab:dset_org} summarises the structure of AUDETER, which comprises two collections: a TTS collection and a Vocoder collection. The TTS collection contains synthetic audio generated by recent end to end TTS systems, with each sample aligned to the script of its corresponding real audio, while the Vocoder collection contains vocoded versions of the same real speech. Both collections are partitioned into four subsets based on the real audio source. For example, the validation subset of the UC subset includes 16,372 real samples and 25 corresponding synthetic variants per sample generated via TTS and vocoding. More details on the usage are discussed in Sec.~\ref{sec:apd:usage}.

\subsubsection{Sources of Real Speech.}
We include four real speech corpora to capture a range of speaking styles and recording conditions, and name their corresponding AUDETER subsets based on their defining characteristics and provenance, namely {\it Celebrity, CrowdSource, US Congress and Audiobook}. The Celebrity, CrowdSource, US Congress, and Audiobook subsets respectively represent in-the-wild public figure speech, user-contributed scripted speech with diverse accents, formal US-accented congressional recordings, and clean read-book speech. We detail the source corpora in Sec. \ref{sec:apd:real}

\subsubsection{Speech Models for Audio Synthesis.}
Table~\ref{tab:syn_models} summarises the speech models used for synthetic audio generation, including 11 open source TTS systems and 10 vocoders, selected for their recency and popularity. For OpenVoice V2, five default speakers are used to produce five synthetic variants to study the effect of voice reference.

\subsubsection{Visualisation of AUDETER's Diversity.} Figures \ref{fig:umap_real} and \ref{fig:umap_fake} visualise our generated deepfake audio samples compared to existing datasets. Both our real and fake samples achieved significantly more diverse coverage.

\subsection{Synthetic Audio Generation Process}
AUDETER includes two synthetic audio generation pipelines: text to waveform synthesis using TTS systems and voice to voice conversion using vocoders. For TTS synthesis, waveform generation uses original transcripts when available, or transcripts obtained via Whisper \cite{whisper} otherwise. For the vocoder collection, selected vocoders are applied directly to real audio samples. The two collections exhibit distinct synthetic patterns. TTS systems encode speaker references and semantic priors from model pretraining, often leveraging large language models, whereas vocoders reflect their pretraining characteristics and artefacts inherited from the original audio. AUDETER’s audio generation pipeline is illustrated in Fig.~\ref{fig:gen_pipeline}.

\subsection{Data Quality Assessment}
To ensure the quality of the generated audio in terms of intelligibility and naturalness, we conduct a thorough evaluation of its comprehensibility and perceptual quality.
 
\subsubsection{Intelligibility Assessments}
\begin{figure}
    \caption{WER similarity (a) and MOS score (b) for the US Congress Subset.}   
    \centering
    \includegraphics[width=1.0\columnwidth]{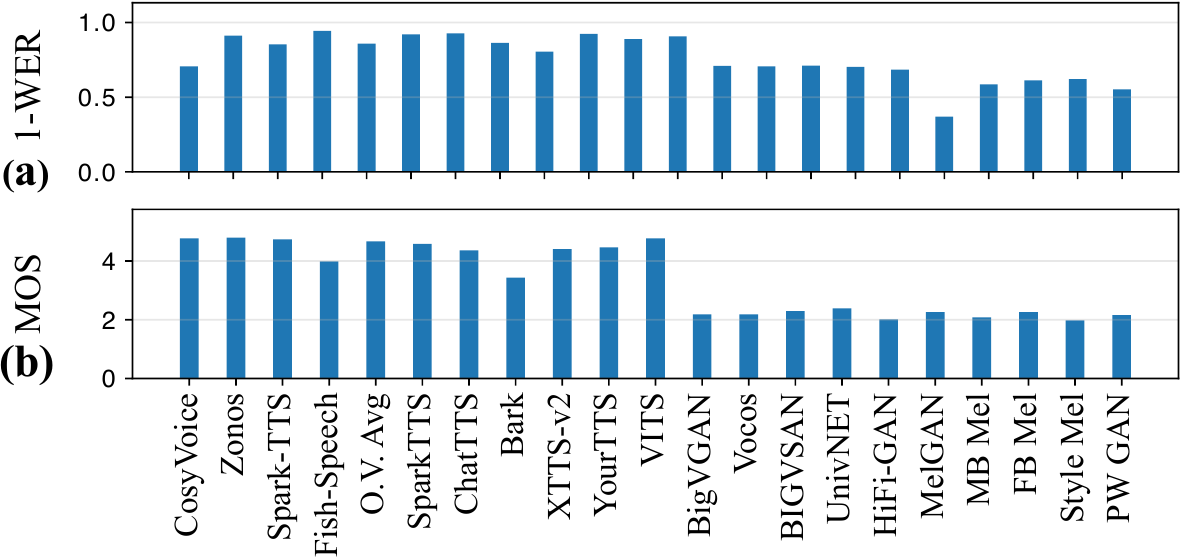}
    \label{fig:intelli}
    \vspace{-0.6cm}
\end{figure}
Intelligibility evaluates whether generated audio preserves the underlying transcript \cite{stevens2005line}. To enable large-scale evaluation, we adopt an automated approach using the Whisper Large V3 ASR model \cite{whisper} to transcribe generated audio and compare it with the reference text. We report four standard metrics: WER similarity (1-WER), word overlap, BLEU, and exact match. Figure~\ref{fig:intelli} (a) shows WER similarity results on the UC subset, with full results for other datasets reported in Section~\ref{apd:intelli}. Overall, TTS models achieve comparable or better intelligibility than the original audio and consistently outperform vocoders, validating the quality of AUDETER’s synthetic samples.

\subsubsection{Naturalness Assessment}
Naturalness evaluates the perceptual similarity of generated audio to human speech. We use Mean Opinion Score (MOS), a standard subjective metric ranging from 1 to 5, and adopt the NISQA framework \cite{nisqa} for automated MOS prediction due to the scale of the dataset. Figure~\ref{fig:intelli} (b) reports average MOS scores across datasets. Similar to intelligibility, modern TTS systems substantially outperform vocoders in naturalness, with multiple models approaching human level perceptual quality.

\section{Experiments}
\begin{figure*}[t]
   \centering
    \caption{Open-world detection performance of the baseline methods on CrowdSource (CS) and US Congress (UC) subsets. }
   \includegraphics[width=1.0\textwidth]{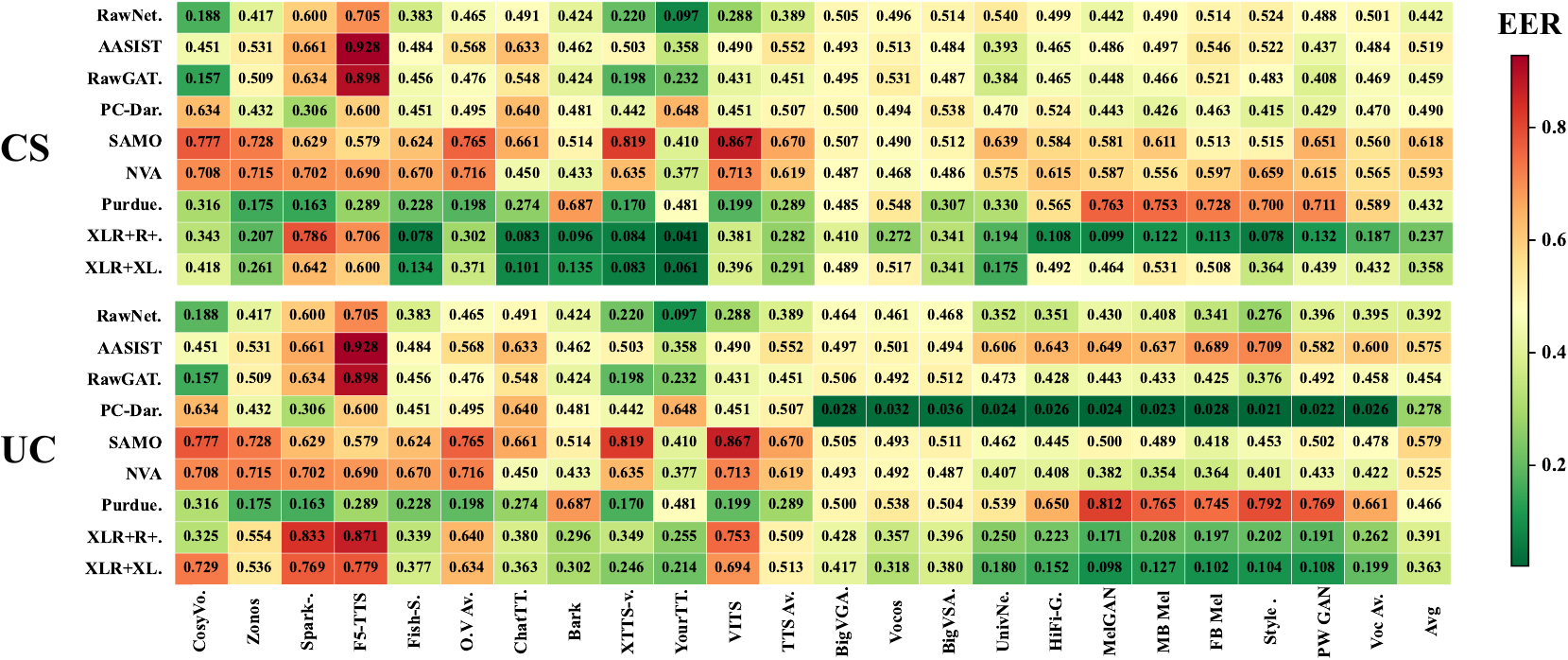}
   \label{tab:main}
   \vspace{-0.5cm}
\end{figure*}
\subsection{Experimental Settings}
\subsubsection{Experiments Overview}
Our experiments are organised into three main parts: evaluating pretrained detectors on AUDETER as a test benchmark (Sec. \ref{sec:testdata}), training existing models on AUDETER for cross-domain evaluation on In-the-Wild and across AUDETER subsets (Sec. \ref{sec:training}), and extending cross-domain analysis to DFADD and SONAR to examine learning under increasingly diverse deepfake sources, motivating and evaluating our proposed curriculum learning strategy (Sec. \ref{sec:cur}).

\subsubsection{Baseline Methods}
We employ nine popular deepfake audio detection methods with publicly available implementations, including both crafted DNN models and detection models that integrate pretrained backbones with scoring heads. Among the first category, we include RawNet2 \cite{rawnet2}, RawGAT-ST \cite{rawgat_st}, AASIST\cite{aasist}, PC-Dart \cite{pc_dart}, SAMO \cite{samo}, Neural Vocoder Artifacts (NVA) \cite{cvprw} and Purdue M2 \cite{m2}. For the second category, we adapt XLS-R + RawNet + Assist (XLS+R+A) \cite{wra} and XLS+SLS \cite{mm}.

\subsubsection{Evaluation Metrics}
Following \cite{rawnet2, itw, aasist, wra, mm}, we employ the popular evaluation metric, Equal Error Rate (EER), to evaluate the detection performance (detailed in Section \ref{sec:ps}). 

\subsubsection{Computational Cost}
Given the scale of AUDETER and varying system efficiency, we estimate our GPU usage. Dataset generation on NVIDIA A100 and H100 clusters consumed approximately 2000 GPU hours for synthetic audio and transcript generation.
 
\subsubsection{Implementation Details}
Pretrained detectors are evaluated using their official implementations and released checkpoints. For supervised training on AUDETER, all models are trained with a learning rate of $1 \times 10^{-6}$ and batch size 128, using a balanced batching strategy with 64 real samples per epoch and 64 uniformly sampled fake samples per batch. All other settings follow the original setup. Importantly, all detection models operate solely on audio inputs and do not access textual transcripts or exploit script matching during either training or inference. For the proposed curriculum learning, we use the same settings for Stage 1 training. Due to space limitations, further details are provided in Sec.\ref{sec:apd:imp}.

\subsubsection{Evaluation Protocol}
\label{sec:eval_protocol}
Since AUDETER contains multiple synthetic versions per real subset, evaluation is performed by iteratively pairing real audio with one synthetic version at a time and averaging performance across runs, as detailed in Algorithm~\ref{alg:eval_eer}. For example, evaluating the CrowdSource subset yields 26 runs across all TTS and vocoder variants. The averaging can be applied per system, per collection, or across the full subset.

\subsection{AUDETER as Test Data} \label{sec:testdata}

We evaluate pretrained baseline models on AUDETER to assess their ability to generalise to novel deepfake patterns and diverse human speech. Figure~\ref{tab:main} summarises performance across speech systems, with complete results on the CE and AB subsets reported in Tables~\ref{tab:apd:itwtts}–\ref{tab:apd:mlsvoc} in the appendix. Overall, baseline methods struggle to achieve robust performance across most settings.

\medskip

\noindent \textbf{Performance on AUDETER.} The baseline models exhibit substantial performance degradation under zero-shot detection on AUDETER, which contains audio generated by recent speech synthesis systems largely absent from existing datasets. No single model achieves consistently low EER across datasets, and the average EER of each model remains high, indicating that generalisation learned from existing benchmarks is insufficient for recent speech systems.

We observe several key trends: (i) detection accuracy drops notably on more recent TTS systems, suggesting increasing acoustic divergence with advances in speech synthesis; (ii) domain shifts in real speech further affect performance, as detectors trained on the same TTS system vary across different real speech corpora; and (iii) models with pretrained backbones achieve relatively stronger overall performance, likely due to improved modelling of real speech, while baseline methods generally perform better on vocoder generated audio. These observations confirm that AUDETER poses meaningful challenges to existing detectors and highlight the importance of evaluating robustness under diverse domain shifts.

\medskip

\noindent \textbf{Effect of Speakers on Detectability.}
To analyse speaker effects on detection performance, we report baseline EERs on synthetic audio generated by five OpenVoice variants across all datasets in Table~\ref{tab:ov_speaker} in the appendix due to space limits. The observed variation across speakers indicates that vocal characteristics influence synthetic audio detectability beyond architectural differences.

\subsection{AUDETER for Training Existing Models}
\label{sec:training}

To demonstrate AUDETER as a useful training resource for open-world detection, we conduct cross-domain evaluations using two popular XLR-based models and one DNN-based detector under their original architectures and training settings. AUDETER-trained models are evaluated on a standard benchmark (ITW) and across different subsets within AUDETER to assess how the training data influences generalisation under various domain shifts.

\begin{table}[t]
\caption{Performance comparison between XLS-R based models trained using our AUDETER dataset with other baseline methods for the in-the-wild (ITW) dataset in EER (\%).}
\begin{adjustbox}{width=0.7 \columnwidth}
\begin{tabular}{cc} \toprule
Model  & EER \\ \midrule
Purdue M2 \cite{m2}                              & 79.75  \\
PC-Dart \cite{pc_dart}                                & 66.17  \\
RawGAT-ST   \cite{rawgat_st}                            & 52.60  \\
AASIST \cite{aasist}                                  & 43.02  \\
RawNet2 \cite{rawnet2}                                & 37.81  \\
SAMO \cite{samo}                                    & 37.09  \\
Wav2vec,HuBERT,Conformer    \& attention \cite{wang2023detection} & 36.84  \\
XLS-R \& Res2Net \cite{yi2023audio}                       & 36.62  \\
MPE \& SENet \cite{wang2024multi}                           & 29.62  \\
NVA \cite{cvprw}                              & 26.32 \\
Spec \& POI-Forensics \cite{pianese2022deepfake}                   & 25.14  \\
XLXS-R,WavLM,Hubert \&   Fusion \cite{yang2024robust}        & 24.27  \\
XLR+R+A \cite{wra}                               & 10.46  \\
XLR-SLS \cite{mm}                               & 7.46   \\ \midrule
XLR+R+A (AUDETER (CS, UC), no aug)    & 5.05   \\
XLR+SLS (AUDETER (CS, UC), no aug)    & 4.17   \\ 
RawNet2 (AUDETER (CS, UC))            & 27.13  \\ 
XLR+R+A (AUDETER (CS, UC))           & 2.03   \\
XLR+SLS (AUDETER (CS, UC))            & 1.87   \\ \bottomrule
\end{tabular}
\end{adjustbox}
\label{tab:itw}
\end{table}

\subsubsection{Cross Domain Generalisation from AUDETER to In-the-Wild}
ITW is the default benchmark for cross-domain DFAD and has no overlap with AUDETER in either real speech sources or synthesis methods. We train XLR+R+A, XLR+SLS, and RawNet2 using  AUDETER’s CS and UC subsets.
Table~\ref{tab:itw} compares the best performance of AUDETER-trained models with baseline DFAD methods. We use reported results when available and otherwise evaluate using the official released weights. AUDETER-trained achieve lower EER than baseline methods and substantially outperform their official pretrained versions. Specifically, EER is reduced from 7.46 to 1.87 for XLR+SLS, from 10.46 to 2.03 for XLR+R+A, and from 37.81 to 27.13 for RawNet2, corresponding to relative reductions of 74.9\%, 80.6\%, and 28.3\%, respectively. Even without the default data augmentation used by existing methods, AUDETER-trained XLR-based models achieve EERs of 5.05 (XLR+SLS) and 4.17 (XLR+R+A), yielding relative reductions of 32.3\% and 60.1\% over their pretrained baselines. This shows that AUDETER improves detection training through diverse real and deepfake patterns, without relying on test-domain knowledge even without data augmentation.

\begin{table}[t]
\caption{Cross-domain DFAD performance on AB subset under only human voice domain shift (All col.) and both domain shift with unseen speech synthesis systems (Unseen col.).}
\begin{adjustbox}{width=0.75\columnwidth}
\begin{tabular}{ccccccc} \toprule
                   & \multicolumn{3}{c}{\textbf{ALL}}           & \multicolumn{3}{c}{\textbf{Unseen}}        \\ \cmidrule(lr){2-4} \cmidrule(lr){5-7} 
{Model}     & {TTS} & {VOC} & {All} & {TTS} & {VOC} & {All} \\ \midrule
{RawNet2}   & 0.257        & 0.434        & 0.346        & 0.286        & 0.378        & 0.332        \\
{AASIST}    & 0.325        & 0.454        & 0.390        & 0.401        & 0.622        & 0.512        \\
{RawGAT-ST} & 0.266        & 0.475        & 0.371        & 0.346        & 0.447        & 0.397        \\
{PC-Dart}   & 0.169        & 0.192        & 0.180        & 0.218        & 0.026        & 0.122        \\
{SAMO}      & 0.492        & 0.506        & 0.499        & 0.471        & 0.475        & 0.473        \\
{NVA} & 0.492        & 0.551        & 0.522        & 0.450        & 0.408        & 0.429        \\
{Purdue-M2} & 0.438        & 0.564        & 0.501        & 0.537        & 0.715        & 0.626        \\
{XLR+R+A}   & 0.365        & 0.192        & 0.279        & 0.346        & 0.239        & 0.292        \\
{XLR+SLS}   & 0.299        & 0.148        & 0.224        & 0.275        & 0.164        & 0.220        \\ \midrule
{XLR+R+A}   & 0.035        & 0.018        & 0.026        & 0.028        & 0.015        & 0.021        \\
{XLR+SLS}   & 0.016        & 0.013        & 0.014        & 0.010        & 0.026        & 0.018        \\
{RawNet 2}  & 0.087        & 0.232        & 0.159        & 0.314        & 0.254        & 0.284     \\ \bottomrule
\end{tabular}
\end{adjustbox}
\label{tab:cd_mls}
\end{table}

\subsubsection{Cross-Domain Generalisation Analysis within AUDETER}
Cross-domain evaluation can be conducted within AUDETER by training on one or more subsets and testing on others. We evaluate generalisation on the AB subset using the three detectors trained on the CS and UC subsets under two settings: (i) training with all speech models' variants, and (ii) training with variants from five selected TTS systems and five vocoders. The first setting isolates human speech domain shift, while the second additionally introduces unseen synthesis systems. Average performance for each setting is reported in Table~\ref{tab:cd_mls}. All three models trained on AUDETER achieve substantial EER reductions compared to their pretrained counterparts, demonstrating improved generalisation under domain shift, highlighting that AUDETER provides diverse and sufficiently rich training data for learning generalisable representations. Among them, XLR-based models achieve near zero EER and consistently outperform RawNet2, showing that sufficient learning capacity from large backbones is essential for encoding generalisable representations.

\subsubsection{Robustness against Real-world Adversaries}
Real-world factors such as background noise can act as adversarial perturbations that impact detection performance. We therefore evaluate the robustness of AUDETER under three widely observed noise conditions: white noise ($\Delta_1$), telephone effect ($\Delta_2$), and background noise ($\Delta_3$). Table~\ref{tab:adv} reports the performance differences ($\Delta$) between evaluations with and without noise perturbations, where smaller values indicate stronger robustness.
We observe that real-world adversaries generally increase EER across models, though the effect is mild. In comparison, models trained on AUDETER exhibit substantially smaller EER increases, demonstrating stronger robustness to these common perturbations. 

\begin{table}[t]
\centering
\caption{Performance difference ($\Delta$ in EER (\%)) between evaluation with and without noise perturbations. Lower values indicate stronger robustness.}
\vspace{-0.3cm}
\label{tab:adv}
\begin{adjustbox}{max width=0.7\columnwidth}
\begin{tabular}{lcccc}
\toprule
\textbf{Model} & $\boldsymbol{\Delta_1}$ & $\boldsymbol{\Delta_2}$ & $\boldsymbol{\Delta_3}$ & $\boldsymbol{\Delta_{\text{avg}}}$ \\
\midrule
XLR+R+A (ori.)              & 1.53 & -1.10 & 14.82 & 5.08 \\
XLR+SLS (ori.)              & 3.23 &  7.33 &  2.20 & 4.25 \\
\midrule
XLR+R+A (AUDE. CS+UC)       & 0.26 &  2.09 &  0.17 & 0.84 \\
XLR+SLS (AUDE. CS+UC)       & 0.28 &  3.77 &  0.29 & 1.45 \\
\bottomrule
\end{tabular}
\end{adjustbox}
\end{table}
\begin{table}[t]
\caption{Performance comparison of two XLR-based detection architectures trained using different combinations of data from AUDETER in EER (\%).}
\vspace{-0.3cm}
\begin{adjustbox}{max width=0.7\columnwidth}
\begin{tabular}{cccc}
\toprule
Collection                     & Subset                   & Model  & Best EER \\ \midrule
\multirow{2}{*}{TTS (All) }           & \multirow{2}{*}{CS+UC}   & XLR+R+A & 11.39          \\
                               &                          & XLR-SLS     &11.59          \\ \cmidrule(lr){2-4}
\multirow{2}{*}{Voc (All) }       & \multirow{2}{*}{CS+UC}   & XLR+R+A  &6.18          \\
                               &                          & XLR-SLS       &6.09          \\ \cmidrule(lr){2-4}
\multirow{2}{*}{TTS + Voc (All)} & \multirow{2}{*}{CS}      & XLR+R+A  &23.71          \\
                               &                          & XLR-SLS       &12.13          \\ \cmidrule(lr){2-4}
\multirow{2}{*}{5 TTS + 5 Voc } & \multirow{2}{*}{CS+UC}     & XLR+R+A &5.52         \\
                               &                          & XLR-SLS        &5.18          \\ \cmidrule(lr){1-4}
\multirow{2}{*}{TTS + Vocoder (All)} & \multirow{2}{*}{CS+UC} & XLR+R+A  & 5.05          \\
                               &                          & XLR-SLS       & 4.17         \\ \bottomrule
\end{tabular}
\end{adjustbox}
\label{tab:diversity}
\end{table}

\subsubsection{Effect of Synthetic Pattern Diversity in Large-scale Training}
To show the benefit of including diverse audio patterns in large-scale training, we vary the subsets to train multiple models and compare their performance on the ITW dataset, as described in Table \ref{tab:diversity}. Specifically, we train the two architecture using: (1) both TTS and Vocoder collections with only the CrowdSource subset, (2) only TTS collection with CrowdSource and US Congress, and (3) only Vocoder collection with CrowdSource and US Congress, using the same settings. For both models, it is not surprising that the models trained using both TTS and Vocoder Collections with samples from both CrowdSource and US Congress subsets yield the best performance, again demonstrating the benefits of training data diversity on open-world detection performance. In contrast, we notice that some models trained with only the CrowdSource subset, even using all TTS and vocoder fake audios, produce unusable performance, further highlighting the usefulness of having diverse real audio in training. 
Although models trained using the CrowdSource and US Congress subsets from either the TTS or the vocoder partition achieve reasonable performance, this is less than when used in combination.

\subsubsection{Analysis of Training Data Scale.}
To show that training with diverse data samples at greater scale can lead to better generalisation, we train the XLR+R+A model using different percentages of all data samples from the CS and UC subsets (10\%, 20\%, 40\%, 60\%, and 100\%) and report the average performance of the top 3 best EER results on the In-the-Wild dataset. We observe consistent improvement as more data are used for training.
\begin{figure}
    \centering
        \caption{Average performance of top 3 EER XLR+R+A models trained with varying data proportions.}

    \includegraphics[width=0.4\linewidth]{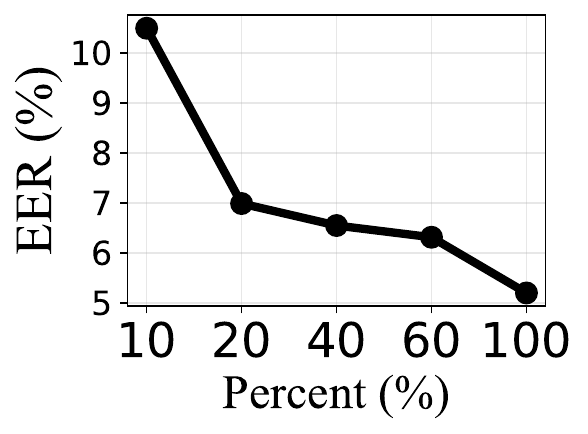}
    \label{fig:enter-label}
\vspace{-0.5cm}
\end{figure}

\subsubsection{Single System Generalisation Test.} \label{sec:single}
Understanding shared characteristics across synthesis models and their impact on cross-system generalisation is important for developing open-world DFAD models. The balanced and consistent structure of AUDETER facilitates this exploration. We train multiple XLR+R+A models using real audio from the CS subset and synthetic audio generated by a single speech synthesis system with matched scripts. Specifically, we consider five TTS models (Fish-Speech, F5-TTS, SparkTTS, VITS, XTTS) and five vocoders (BigVGAN, UnivNet, MelGAN, HiFi-GAN, Vocos), repeating the training process ten times for each system. Evaluation follows the protocol in Sec.\ref{sec:eval_protocol}. Due to space constraints, Table~\ref{tab:single_sys} reports average performance across the TTS and Vocoder collections over three test domains, with full results provided in Tables~\ref{tab:apd24}–\ref{tab:apd35}.

\bigskip

\noindent \textbf{Generalisation on Matched Synthetic Audio and Identical Human Speech.}
The results in the CS Val column reflect average performance of models trained on single systems within each collection, where scripts and real audio match the training data, isolating the effect of synthetic patterns. Generalisation varies markedly across speech models, with SparkTTS and Fish-Speech exhibiting stronger transfer among TTS systems. Cross-collection generalisation between TTS systems and vocoders remains limited, with models performing better within the same collection, and vocoders generally exhibiting weaker generalisation. These results highlight the importance of including both system types and diverse synthesis models for robust detector training.

\smallskip

\noindent \textbf{Generalisation with Different Textual Content.}
We repeat the evaluation on the CS test partition, where textual content differs from training while the same speaker style is preserved. Results in the CS Test column are slightly lower but comparable to those with matched text, indicating that textual content has limited impact on generalisation.

\smallskip

\noindent \textbf{Generalisation across Domains and Text.} We further evaluate on the UC test partition to assess the effect of distribution shift in real speech. Performance drops substantially compared to previous settings, even when using the same synthesis system as in training, indicating high sensitivity to real-audio domain shift and underscoring the importance of diverse real speech in training data.

\begin{table}[t]
\caption{Single system generalisation performance of the three detection models in EER (\%).}
\begin{adjustbox}{max width=\columnwidth}
\begin{tabular}{ccccccc} \toprule
\multicolumn{1}{c}{} & \multicolumn{2}{c}{CS Train} & \multicolumn{2}{c}{CS Test} & \multicolumn{2}{c}{UC Test} \\ \cmidrule(lr){2-3} \cmidrule(lr){4-5} \cmidrule(lr){6-7}
Single Sys.          & TTS AVG      & Voc Avg     & TTS AVG      & Voc Avg      & TTS AVG      & Voc Avg      \\  \midrule
SparkTTS             & 9.6        & 49.6       &10.0        &49.6        & 48.2        & 54.1        \\
F5-TTS               & 15.0       & 49.2       &15.5        &49.1        &66.0        &56.9        \\
Fish-Speech          & 9.3        & 43.8       &9.7        &44.2        &51.0        & 46.4        \\
XTTS                 & 17.2        &49.6       &17.8        &49.6        &66.9        &53.1        \\
VITS                 & 10.8        &48.5       &11.1        &48.5        &45.5        &53.3        \\
BigVGAN              & 54.7        &27.0       &56.5        &29.6        &49.5        &42.6        \\
HiFi-GAN             & 23.8        &28.5       &24.9        &29.0        &69.1        &44.1        \\
Vocos                & 26.2        &6.2       &30.2        &9.0        & 61.3        & 32.0        \\
UnivNet              &23.3      &28.5         &21.2        &12.6        & 39.9        &22.7        \\
Mel GAN              & 23.5        &19.9       & 24.5        & 20.1        &69.7        & 31.5        \\ \bottomrule
\end{tabular}
\end{adjustbox}
\label{tab:single_sys}
\vspace{-0.5cm}
\end{table}
\subsection{Towards Diverse Deepfake Training} \label{sec:cur}
\subsubsection{Limitations of Binary Classification for Training Detection Models}
Regardless of model architecture, most existing DFAD models are trained as a binary classification problem. Beyond limited generalisation to unseen systems, this formulation can induce negative transfer when trained on highly diverse deepfake sources. In particular, certain speech synthesis models exhibit strong system specific fingerprints, which can bias the training signal and encourage models to overfit to these cues, limiting the learning of representations that generalise across most systems. We refer to such systems as harmful systems. 
We observe that including such harmful systems under current training schemes can substantially degrade cross domain performance. To better study this effect, we incorporate two recent evaluation focused datasets: DFADD and SONAR, alongside ITW to strengthen cross-domain evaluation.

\begin{figure}[t]
   \centering
   \vspace{-0.3cm}
    \caption{Linear probing visualisation of system specific fingerprint dominance, with Spark TTS and BigVGAN most pronounced (Spark. and BigVGA. in the plot).} 
   \includegraphics[width=\columnwidth]{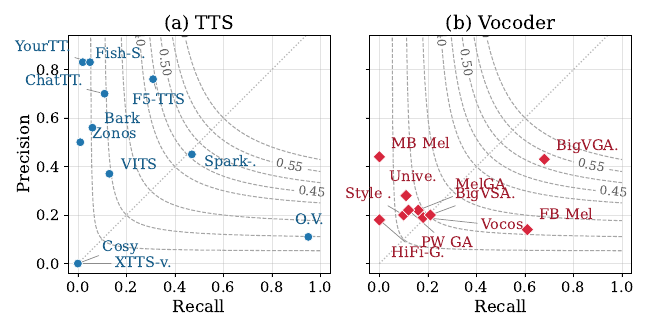}
   \label{fir:probing}
   \vspace{-0.5cm}
\end{figure}

\begin{figure}[htbp]
   \centering
   \vspace{-0.4cm}
    \caption{Our proposed curriculum learning method, where C and H denotes core and harmful systems, respectively.}
   \includegraphics[width=0.9\columnwidth]{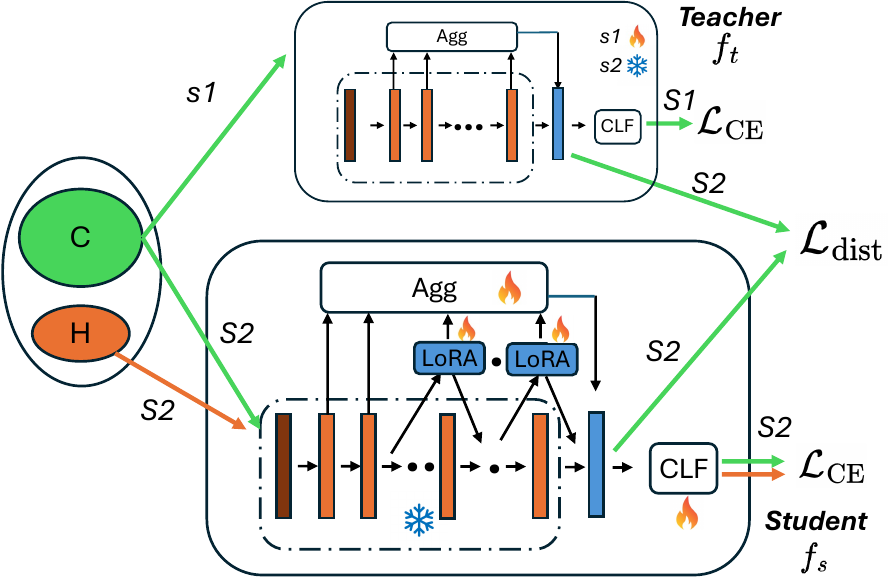}
   \label{fig:cur}
   \vspace{-0.3cm}
\end{figure}

\subsubsection{Proposed Baseline Method} As a complementary contribution, we propose a two stage curriculum training strategy to mitigate negative transfer caused by harmful systems (Fig.~\ref{fir:probing}). The core intuition is to first establish system-invariant and universal representations using data samples generated by systems that exhibit weaker system specific artefacts, before gradually incorporating those harmful system with more pronounced fingerprints.

The first stage involves two main tasks: identifying harmful systems via linear probing and training a base detector using system that are considered as non-harmful, which we refer to as core-systems. The linear probing test aims to identify systems that expose strong and persistent system specific cues in the representations learned for DFAD. We freeze the backbone of a detector trained on all systems and optimise a lightweight system identification head as a proxy for system harmfulness. After light tuning of the head, harmful systems exhibit balanced precision and recall, indicating persistent system-specific cues that signal early dominance. Figure \ref{fig:cur} shows the results of the probing test on AUDETER systems. For our baseline model, we treat the most dominant systems from each subset: SparkTTS for TTS and BigVGAN for vocoders, as harmful systems and use the remaining systems to train a base detector.

In the second stage, we adopt a student–teacher paradigm as a regulariser for supervised training on the full system set. The motivation is to prevent harmful systems from overshadowing the system-invariant cues learned in Stage 1. Strongly dominant systems can exert a disproportionate influence during training, and even when they constitute only a small fraction of the data, they may bias the representation toward system-specific artefacts rather than generalisable cues. Let $f_t$ and $f_s$ denote the teacher (Stage-1) and student models, respectively, producing logits $\mathbf{z}_t, \mathbf{z}_s \in \mathbb{R}^2$ for an input sample $x$; $\mathcal{D}_{\text{core}}$ denote the core system set and $\mathcal{D}_{\text{full}}$ the full training set, then in Stage~2, we fix $f_t$ and optimise $f_s$ by minimising
\begin{equation}
\mathcal{L}
=
\mathcal{L}_{\mathrm{CE}}(\mathbf{z}_{s}, y)
+
\lambda(x)\,
\mathcal{L}_{\mathrm{dist}}(\mathbf{z}_{t}, \mathbf{z}_{s}),
\end{equation}
where $\lambda(x)=\lambda$ if $x \in \mathcal{D}_{\text{core}}$ and $\lambda(x)=0$ otherwise. The distillation loss is defined as:
\begin{equation}
\mathcal{L}_{\mathrm{dist}}
=
\mathrm{KL}\!\left(
\mathrm{softmax}(\mathbf{z}_{t})
\,\|\, 
\mathrm{softmax}(\mathbf{z}_{s})
\right).
\end{equation}
This encourages the student model to preserve the generalisable cues learned by the teacher while adapting to newly introduced systems. To further improve training stability, we employ LoRA adapters in the last three layers and freeze all other parameters, unfreezing only the aggregator module to learn multi-layer representations. During inference, the student model is used end-to-end with the LoRA weights merged for scoring, introducing negligible overhead to the base architecture. Framework details are presented in Section~\ref{sec:apd.imp.cur}.

\subsubsection{Cross-dataset Performance}
\begin{table}[t]
\centering
\caption{Performance comparison on the extended evaluation set between the proposed curriculum learning method (cur), the AUDETER-trained supervised model (sup), and pretrained baseline detectors.}
\vspace{-0.3cm}
\label{tab:curr}
\begin{adjustbox}{width=0.9\columnwidth} 
\begin{tabular}{lcccc cc}
\toprule
 & \multicolumn{4}{c}{\textbf{Cross Domain}} & \textbf{In Domain}  &\textbf{Overall}\\
\cmidrule(lr){2-5} \cmidrule(lr){6-6} \cmidrule(lr){7-7}
\textbf{Model} & ITW & DFAFF & SONAR & Avg &AUDETER  & Avg \\
\midrule
{RawNet2}   &37.81	&41.86	&43.00	&40.89	&44.20	&32.80   \\
{AASIST}    &43.02	&31.86	&57.47	&44.12	&51.90	&45.67   \\
{RawGAT-ST} &52.60	&23.70	&50.78	&42.36	&23.70	&38.63   \\
{PC-Dart}   &66.17	&85.17	&51.90	&67.75  &38.40	&60.41 \\
{SAMO}      &37.09	&78.68	&42.08	&52.62  &59.85	&54.43  \\
{NVA}       &26.32	&51.97	&74.61	&50.97  &55.90	&52.20  \\
{Purdue-M2} &79.57	&58.41	&73.00	&70.33  &44.90	&63.97   \\
XLR+R+A     &10.46	&11.93	&46.12	&22.84	&23.70	&23.05   \\
XLR+SLS     &7.46	&7.54	&24.72	&13.24	&35.80	&18.88    \\
\midrule
XLR+R+A (sup)  & 2.03  &27.68 & 13.83 & 14.51 &0.87 &11.10    \\
XLR+SLS (sup)  &\textbf{1.87}  &22.80  &14.58  &13.08  & \textbf{0.71} &9.99  \\
XLR+R+A (cur)  & 2.31  &\textbf{6.27}  &\textbf{12.78} & \textbf{7.12}  &1.48 &\textbf{5.71} \\
\bottomrule
\end{tabular}
\end{adjustbox}
\vspace{-0.4cm}
\end{table}
Table~\ref{tab:curr} summarises the results on the extended benchmark. The proposed method achieves substantially lower EER than both pretrained baselines and AUDETER-trained XLR models using all systems in terms of cross-domain and overall performance. Although a minor in-domain performance drop is observed, this trade-off is worthwhile, as the method avoids overfitting to systems with strong synthesis fingerprints and improves cross-domain generalisation. This demonstrates that the proposed method effectively learns balanced and robust representations, avoiding dominance by any single synthesis system.

\section{Conclusion and Future Works}
This paper presents AUDETER, a large-scale deepfake audio detection dataset for enhanced evaluation and training generalised detection models for open-world detection. It contains nearly 3 million synthetic audio generated by us using 21 recent speech synthesis systems paired with diverse real speech from 4 corpora, enabling controlled, fine-grained evaluation under various domain shifts. Extensive empirical evaluation shows that AUDETER effectively challenges existing detection models, and that training on AUDETER’s large scale and diverse synthetic patterns leads to significantly improved generalisation. As a complementary contribution, we identify that the default training schemes used by existing methods suffer from negative transfer when trained on diverse deepfake data, and we propose a curriculum learning strategy to mitigate this effect. Looking forward, we view AUDETER as a foundation for developing more effective training paradigms for deepfake detection. In particular, future work will explore self supervised and representation learning approaches tailored to deepfake audio, with the goal of learning robust, system invariant features from diverse training data.

\bibliographystyle{ACM-Reference-Format}
\bibliography{ref}

\onecolumn
\appendix
\section{Detailed Dataset Information}
\FloatBarrier
\subsection{Dataset Generation Pipeline}
\begin{figure}[h]
    \centering
    \includegraphics[width=0.8\columnwidth]{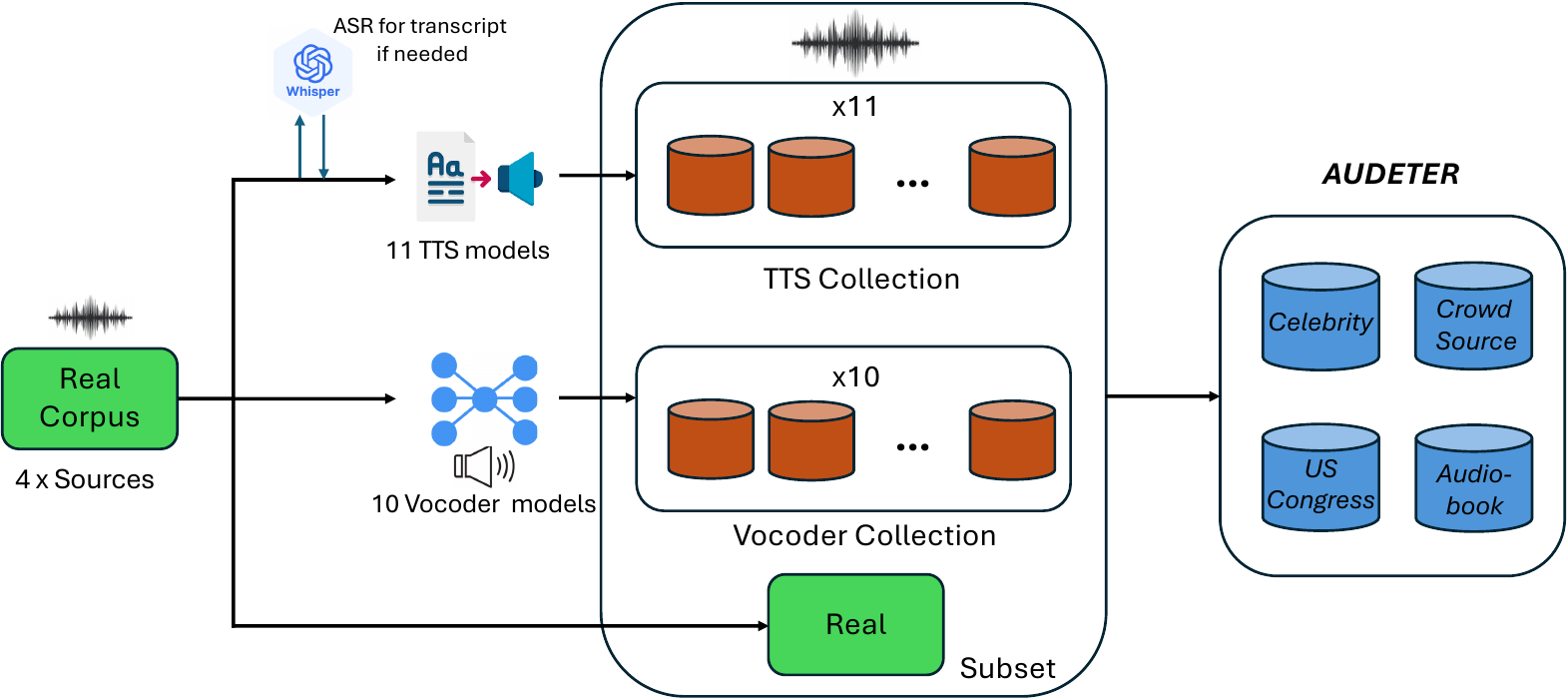}
    \caption{Illustration of AUDETER’s data generation pipeline.}
    \label{fig:gen_pipeline}
\end{figure}

\subsection{Audio Synthesis Models}
We employ both recent end-to-end TTS systems and legacy vocoders for synthetic audio generation. We provide their descriptions in this section.

\begin{table}[h]
\centering
\label{tab:tts_models}
\begin{tabular}{cc} \toprule

\textbf{Model} & \textbf{Link} \\  \midrule

CosyVoice & \url{https://github.com/FunAudioLLM/CosyVoice} \\ 

Zonos & \url{https://github.com/Zyphra/Zonos} \\

Spark-TTS & \url{https://github.com/SparkAudio/Spark-TTS} \\

F5-TTS & \url{https://github.com/SWivid/F5-TTS} \\

Fish-Speech & \url{https://github.com/fishaudio/fish-speech} \\
OpenVoice & \url{ https://github.com/myshell-ai/OpenVoice} \\
ChatTTS & \url{https://github.com/2noise/ChatTTS} \\

Bark & \url{https://github.com/suno-ai/bark} \\

XTTS-v2 & \url{https://github.com/coqui-ai/TTS} \\

YourTTS & \url{https://github.com/Edresson/YourTTS} \\

VITS & \url{https://github.com/jaywalnut310/vits} \\
\bottomrule
\end{tabular}
\caption{TTS Models and Their GitHub Repositories}
\end{table}

\subsubsection{TTS Models Selection}
We include 11 popular TTS models for text to waveform deepfake audio generation. To investigate the effect of different speaker references, for OpenVoice V2, we generate 5 versions using its default speakers with 5 different English Accent accents, totalling 15 different versions. A summary of the key dataset information are reported in Table.

\subsubsection{Vocoder Models Selection}
We include 10 popular vocoder models for constructing the vocoder collection, and summarise their details in Table \ref{tab:vocl}.

\begin{table}[h]
\begin{tabular}{cc} \toprule
Vocoder    & Link     \\ \midrule
BigVGAN    & { { https://github.com/NVIDIA/BigVGAN}}              \\
Vocos      & { { https://github.com/gemelo-ai/Vocos}}             \\
BigVSAN    & { { https://github.com/sony/bigvsan}}                \\
UnivNet V2 & { { https://github.com/maum-ai/univnet}}             \\
HiFi-GAN   & { { https://github.com/kan-bayashi/ParallelWaveGAN}} \\
MelGAN     & { { https://github.com/kan-bayashi/ParallelWaveGAN}} \\
MB Mel     & { { https://github.com/kan-bayashi/ParallelWaveGAN}} \\
FB Mel     & { { https://github.com/kan-bayashi/ParallelWaveGAN}} \\
Style Mel  & { { https://github.com/kan-bayashi/ParallelWaveGAN}} \\
PW GAN     & { { https://github.com/kan-bayashi/ParallelWaveGAN}} \\ \bottomrule
\end{tabular}
\caption{Link to our selected vocoder models.}
\label{tab:vocl}
\end{table}

\subsection{Real Audio}\label{sec:apd:real}
We summarise the four corpora of of real huamn voice in Table \ref{tab:apd:real}. The real-speech subsets in AUDETER are named based on their speech characteristics and recording conditions rather than the original dataset titles. Each subset is derived from an existing real-speech corpus and used solely as a source of human speech for synthetic audio generation and evaluation:
\begin{itemize}
    \item Celebrity is derived from the In-the-Wild dataset and contains recordings of public figures captured under diverse real-world conditions.
    \item CrowdSource is derived from Common Voice 13.0, consisting of speech uploaded by public users, with read scripts recorded across a wide range of accents, devices, and acoustic environments.
    \item US Congress is derived from The People’s Speech, which includes formal recordings of U.S. congressional sessions and related public-domain speech.
    \item Audiobook is derived from Multilingual LibriSpeech (MLS) and contains clean, professionally recorded read speech from audiobooks.
\end{itemize}

\begin{table}[h] 
\scalebox{0.9}{
\begin{tabular}{cc} \toprule
Dataset   Name                 & Link                                                                                                         \\ \midrule
In-the-Wild                    & https://deepfake-total.com/in\_the\_wild                                                               \\ 
Common Voice 13. 0             & Hugging Face Datasets: mozilla-foundation/common\_voice\_13\_0 (validation and test partition)             \\
The People's Speech            & Hugging Face Datasets: MLCommons/peoples\_speech (train and test   partition from the clean subset)     \\
Multilingual LibriSpeech (MLS) & Hugging Face Datasets: parler-tts/mls\_eng (English version of the   Multilingual LibriSpeech (MLS) dataset) \\ \bottomrule
\end{tabular}
}
\caption{Links to real voice corpora.}
\label{tab:apd:real}
\end{table}

\subsection{Deepfake Audio}\label{sec:apd:usage}
All AUDETER  subsets include predefined train and test partitions, with the exception of Celebrity, which is constructed from our curated version of In-the-Wild data and is primarily used for cross-domain evaluation against the standard ITW benchmark.
\begin{itemize}
    \item For within-AUDETER experiments, models are trained on the train split and evaluated on the test split of each subset.
    \item For cross-domain evaluation (e.g., evaluation on ITW, DFADD, or SONAR), both train and test splits of AUDETER subsets may be used for training to provide the strongest possible training coverage.
    \item The Celebrity subset is not used for within-dataset training and testing splits. Instead, it serves as a controlled AUDETER counterpart to the classic ITW dataset for cross-domain comparison.
\end{itemize}

\subsection{Detailed Dataset Statistics}
We present the detailed hour counts for each partition of the TTS collection in Table \ref{tab-apd:tts_hour} and the vocoder collection in Table \ref{tab-apd:voc_hour}.

\begin{table*}[h]
\begin{tabular}{ccccccccccc}
\toprule
\multicolumn{1}{l}{} & \multicolumn{10}{c}{TTS Collection}                                                                           \\ \cmidrule(lr){2-11}
                     & Celebrity & \multicolumn{3}{c}{CrowdSource} & \multicolumn{3}{c}{US Congress} & \multicolumn{3}{c}{Audiobook}  \\ \cmidrule(lr){2-2} \cmidrule(lr){3-5} \cmidrule(lr){6-8} \cmidrule(lr){9-11}
 Model                    & Test   & Val       & Test      & Total    & Val       & Test      & Total     & Test    & Test   & Total  \\ \midrule
CosyVoice            & 16.25       & 14.76     & 14.00     & 28.76    & 27.15     & 50.04     & 77.19     & 12.48  & 12.45  & 24.93  \\
Zonos                & 18.01       & 16.20     & 15.48     & 31.68    & 30.09     & 55.50     & 85.59     & 12.94  & 12.96  & 25.90  \\
Sparktts             & 18.89       & 18.24     & 17.44     & 35.68    & 30.87     & 56.88     & 87.75     & 13.95  & 13.91  & 27.86  \\
F5-TTS               & 27.91       & 25.58     & 24.49     & 50.07    & 48.83     & 89.83     & 138.66    & 21.75  & 21.68  & 43.43  \\
Fish-Speech          & 27.03       & 18.07     & 18.70     & 36.77    & 40.70     & 74.27     & 114.97    & 33.75  & 30.55  & 64.30  \\
OV2 S1               & 20.74       & 18.92     & 18.20     & 37.12    & 31.16     & 57.62     & 88.78     & 12.79  & 12.77  & 25.56  \\
OV2 S2               & 17.28       & 15.90     & 15.23     & 31.13    & 27.37     & 50.56     & 77.93     & 11.55  & 11.54  & 23.09  \\
OV2 S3               & 19.53       & 17.93     & 17.24     & 35.17    & 30.04     & 55.52     & 85.56     & 12.40  & 12.39  & 24.79  \\
OV2 S4               & 19.98       & 18.02     & 17.36     & 35.38    & 29.98     & 55.39     & 85.37     & 12.07  & 12.06  & 24.13  \\
OV2 S5               & 16.02       & 14.74     & 14.09     & 28.83    & 25.61     & 47.29     & 72.90     & 11.04  & 11.04  & 22.08  \\
ChatTTS              & 24.95       & 21.73     & 20.98     & 42.71    & 44.33     & 81.13     & 125.46    & 20.49  & 20.28  & 40.77  \\
Bark                 & 29.05       & 24.79     & 24.42     & 49.21    & 40.18     & 74.26     & 114.44    & 13.96  & 13.79  & 27.75  \\
XTTS                 & 18.54       & 16.37     & 15.71     & 32.08    & 30.88     & 57.05     & 87.93     & 12.78  & 12.77  & 25.55  \\
YourTTS              & 20.14       & 18.56     & 17.76     & 36.32    & 30.88     & 57.05     & 87.93     & 12.78  & 12.84  & 25.62  \\
VITS                 & 17.16       & 15.21     & 14.17     & 29.38    & 25.42     & 47.01     & 72.43     & 10.49  & 10.52  & 21.01  \\ \midrule
      Total               & 311.48      & 275.02    & 265.27    & 540.29   & 493.49    & 909.40    & 1402.89   & 225.22 & 221.55 & 446.77 \\ \bottomrule
\end{tabular}
\caption{Detailed time information of the TTS collection.}
\label{tab-apd:tts_hour}
\end{table*}
\FloatBarrier

\begin{table*}[h]
\begin{tabular}{ccccccccccc}
\toprule
\multicolumn{1}{l}{} & \multicolumn{10}{c}{TTS Collection}                                                                           \\ \cmidrule(lr){2-11}
                     & Celebrity & \multicolumn{3}{c}{CrowdSource} & \multicolumn{3}{c}{US Congress} & \multicolumn{3}{c}{Audiobook}  \\ \cmidrule(lr){2-2} \cmidrule(lr){3-5} \cmidrule(lr){6-8} \cmidrule(lr){9-11}
Model                & Bona-fide   & Val       & Test      & Total    & Val       & Test      & Total     & Val    & Test   & Total  \\ \midrule
BigVGAN              & 20.73       & 27.24     & 27.04     & 54.28    & 33.14     & 59.75     & 92.89     & 15.75  & 15.54  & 31.29  \\
Vocos                & 20.73       & 27.24     & 27.04     & 54.28    & 33.14     & 59.75     & 92.89     & 15.75  & 15.54  & 31.29  \\
BigVSAN              & 20.73       & 27.16     & 26.96     & 54.12    & 33.14     & 59.75     & 92.89     & 15.76  & 15.55  & 31.31  \\
UnivNet              & 20.73       & 25.88     & 25.71     & 51.59    & 33.14     & 59.75     & 92.89     & 15.42  & 15.42  & 30.84  \\
HiFi-GAN             & 20.79       & 25.88     & 25.71     & 51.59    & 33.19     & 59.86     & 93.05     & 15.59  & 15.42  & 31.01  \\
MelGAN               & 20.77       & 25.88     & 25.71     & 51.59    & 33.19     & 59.85     & 93.04     & 15.59  & 15.42  & 31.01  \\
MB Mel               & 20.77       & 27.16     & 26.96     & 54.12    & 33.19     & 59.85     & 93.04     & 15.76  & 15.55  & 31.31  \\
FB Mel               & 20.77       & 25.88     & 25.71     & 51.59    & 33.19     & 59.85     & 93.04     & 15.59  & 15.42  & 31.01  \\
Style Mel            & 20.79       & 27.24     & 27.04     & 54.28    & 33.19     & 59.86     & 93.05     & 15.75  & 15.54  & 31.29  \\
PW GAN               & 20.77       & 27.11     & 26.91     & 54.02    & 33.19     & 59.85     & 93.04     & 15.75  & 15.54  & 31.29  \\ \midrule
Total                & 207.58      & 266.67    & 264.79    & 531.46   & 331.7     & 598.12    & 929.82    & 156.71 & 154.94 & 311.65 \\ \bottomrule
\end{tabular}
\caption{Detailed time information of the Vocoder collection.}
\label{tab-apd:voc_hour}
\end{table*}
\FloatBarrier

\section{Complete Datasets Quality Assurance Results} \label{apd:intelli}
We provide the detailed results for intelligibility assessment and naturalness.

\begin{table}[h]
\scalebox{0.75}{
\begin{tabular}{ccccccccccccccccc}
\toprule
Metricc      & Original & CosyVoice & Zonos & Spark-TTS & F5-TTS & Fish-Speech & O.V S1 & O.V S2 & O.V S3 & O.V S4 & O.V S5 & ChatTTS & Bark  & XTTS-v2 & YourTTS & VITS  \\\hline
WER          & 0.873    & 0.935     & 0.869 & 0.962     & 0.920  & 0.924       & 0.950  & 0.898  & 0.950  & 0.953  & 0.955  & 0.887   & 0.850 & 0.946   & 0.894   & 0.908 \\
Word Overlap & 0.846    & 0.905     & 0.826 & 0.943     & 0.924  & 0.888       & 0.924  & 0.858  & 0.924  & 0.929  & 0.932  & 0.853   & 0.842 & 0.921   & 0.853   & 0.874 \\
BLEU         & 0.816    & 0.887     & 0.799 & 0.931     & 0.912  & 0.866       & 0.910  & 0.833  & 0.910  & 0.915  & 0.917  & 0.833   & 0.823 & 0.910   & 0.834   & 0.850 \\
Exact Match  & 0.545    & 0.659     & 0.469 & 0.780     & 0.705  & 0.613       & 0.712  & 0.542  & 0.718  & 0.734  & 0.722  & 0.519   & 0.509 & 0.682   & 0.510   & 0.592\\
\bottomrule
\end{tabular}
}
\caption{Intelligibility results of the CrowdSource subset in WER, Word Overlap, BLEU, and Exact Match for TTS models. }
\label{tab:apd:intcs}
\end{table}

\begin{table}[h]
\scalebox{0.75}{
\begin{tabular}{ccccccccccccccccc}
\toprule
Metricc      & Original & CosyVoice & Zonos & Spark-TTS & F5-TTS & Fish-Speech & O.V S1 & O.V S2 & O.V S3 & O.V S4 & O.V S5 & ChatTTS & Bark  & XTTS-v2 & YourTTS & VITS  \\\hline
WER          & 0.707    & 0.912     & 0.853 & 0.943     & 0.859  & 0.921       & 0.931  & 0.906  & 0.910  & 0.944  & 0.944  & 0.863   & 0.805 & 0.924   & 0.888   & 0.907 \\
Word Overlap & 0.765    & 0.891     & 0.827 & 0.931     & 0.912  & 0.895       & 0.906  & 0.880  & 0.886  & 0.924  & 0.925  & 0.848   & 0.832 & 0.913   & 0.864   & 0.879 \\
BLEU         & 0.724    & 0.873     & 0.798 & 0.917     & 0.891  & 0.871       & 0.887  & 0.857  & 0.864  & 0.909  & 0.909  & 0.817   & 0.802 & 0.896   & 0.844   & 0.850 \\
Exact Match  & 0.239    & 0.562     & 0.402 & 0.688     & 0.584  & 0.540       & 0.583  & 0.525  & 0.256  & 0.650  & 0.650  & 0.411   & 0.405 & 0.603   & 0.501   & 0.474\\
\bottomrule
\end{tabular}
}
\caption{Intelligibility results of the US Congress subset in WER, Word Overlap, BLEU, and Exact Match for TTS models.}
\label{tab:apd:intuc}
\end{table}
                                             
\FloatBarrier
\subsection{Detailed Intelligibility Scores}
We provide detailed intelligibility results in Table~\ref{tab:apd:intcs} and Table~\ref{tab:apd:intuc} for the CrowdSource and US Congress subsets, respectively.

\subsection{Detailed Naturalness Scores}
We provide the detailed results for the NISQA naturalness assessmenton all our deepfake audio samples and present the results in Table \ref{tab:apd:mos}.

\begin{table}[h]
\begin{tabular}{cccccccc} \toprule
            & Celebrity       & \multicolumn{2}{c}{CrowdSource} & \multicolumn{2}{c}{US Congress} & \multicolumn{2}{c}{Audiobook} \\ \cmidrule(lr){2-2} \cmidrule(lr){3-4} \cmidrule(lr){5-6}  \cmidrule(lr){7-8} 
            & bona-fide & val        & test      & val        & test      & dev        & test       \\ \midrule
CosyVoice   & 4.75      & 4.817      & 4.81      & 4.781      & 4.771     & 4.806      & 4.805      \\
Zonos       & 4.736     & 4.856      & 4.844     & 4.805      & 4.798     & 4.848      & 4.859      \\
Spark-TTS   & 4.752     & 4.797      & 4.79      & 4.739      & 4.736     & 4.747      & 4.742      \\
F5-TTS      & 3.908     & 4.163      & 4.128     & 3.981      & 3.997     & 4.053      & 4.061      \\
Fish-Speech & 4.353     & 4.379      & 4.371     & 4.423      & 4.41      & 4.491      & 4.493      \\
O.V s1    & 4.466     & 4.506      & 4.51      & 4.548      & 4.534     & 4.619      & 4.639      \\
O.V s2    & 4.566     & 4.622      & 4.619     & 4.53       & 4.524     & 4.537      & 4.534      \\
O.V s3    & 4.977     & 5.007      & 5.007     & 5.005      & 5.002     & 5.011      & 5.008      \\
O.V s4    & 4.825     & 4.894      & 4.89      & 4.851      & 4.85      & 4.849      & 4.857      \\
O.V s5    & 4.521     & 4.636      & 4.633     & 4.589      & 4.577     & 4.611      & 4.6        \\
ChatTTS     & 4.264     & 4.466      & 4.451     & 4.359      & 4.371     & 4.378      & 4.373      \\
Bark        & 3.352     & 3.562      & 3.521     & 3.434      & 3.429     & 3.69       & 3.652      \\
XTTS     & 4.302     & 4.441      & 4.414     & 4.413      & 4.398     & 4.604      & 4.601      \\
YourTTS     & 4.29      & 4.487      & 4.469     & 4.474      & 4.457     & 4.603      & 4.601      \\
VITS        & 4.677     & 4.778      & 0.468         & 4.772      & 4.767     & 4.824      & 4.824      \\
BigVGAN     & 3.287     & 3.105      & 2.945     & 2.081      & 2.291     & 3.694      & 3.805      \\
Vocos       & 3.27      & 3.079      & 2.92      & 2.088      & 2.273     & 3.633      & 3.738      \\
BigVSAN     & 3.267     & 3.118      & 2.955     & 2.19       & 2.406     & 3.731      & 3.819      \\
UnivNet     & 3.359     & 3.437      & 3.27      & 2.34       & 2.443     & 3.849      & 3.902      \\
HiFi-GAN    & 3.014     & 2.995      & 2.865     & 1.952      & 2.074     & 3.45       & 3.572      \\
MelGAN      & 2.987     & 2.941      & 2.801     & 2.226      & 2.3       & 3.221      & 3.252      \\
MB Mel      & 2.874     & 3.043      & 2.92      & 2.052      & 2.116     & 3.393      & 3.383      \\
FB Mel      & 3.137     & 3.267      & 3.134     & 2.22       & 2.299     & 3.634      & 3.626      \\
Style Mel   & 2.826     & 2.913      & 2.801     & 1.922      & 2.036     & 3.35       & 3.421      \\
PW GAN      & 2.977     & 3.193      & 3.062     & 2.082      & 2.226     & 3.547      & 3.541     \\ \bottomrule
\end{tabular}
\caption{The MOS scores on all subsets.}
\label{tab:apd:mos}
\end{table}

\section{Implementation Details}\label{sec:apd:imp}
\subsection{EER Evaluation with Multiple Synthetic Variants}
Pseudocode for the evaluation protocol with multiple synthetic variants is provided in Algorithm~\ref{alg:eval_eer}, 
\begin{algorithm}[t]
\caption{EER Evaluation with Multiple Synthetic Variants}
\label{alg:eval_eer}
\begin{algorithmic}[1]
\Require Trained detector $f$, real audio set $\mathcal{R}$, synthetic variants $\{\mathcal{S}_k\}_{k=1}^{K}$
\Ensure Averaged Equal Error Rate $\overline{\mathrm{EER}}$

\State $\overline{\mathrm{EER}} \leftarrow 0$
\For{$k = 1$ to $K$}
    \State Construct evaluation set $\mathcal{E}_k = \mathcal{R} \cup \mathcal{S}_k$
    \State Compute $\mathrm{EER}_k$ by evaluating $f$ on $\mathcal{E}_k$
    \State $\overline{\mathrm{EER}} \leftarrow \overline{\mathrm{EER}} + \mathrm{EER}_k$
\EndFor
\State $\overline{\mathrm{EER}} \leftarrow \overline{\mathrm{EER}} / K$
\State \Return $\overline{\mathrm{EER}}$
\end{algorithmic}
\end{algorithm}

\subsection{Curriculum Learning}\label{sec:apd.imp.cur}
We adopt a two-stage curriculum learning strategy to incorporate unseen TTS systems while preserving performance on previously learned core systems. Stage 1 trains the detector on 19 core TTS methods using standard cross-entropy with RawBoost augmentation. Stage 2 introduces two new systems (SparkTTS and BigVGAN) via teacher–student distillation with LoRA, where rank-4 adapters ($\alpha=8$) are inserted into the Q, K, V, and output projections of XLS-R transformer layers 21–23, with Kaiming-initialised A and zero-initialised B to ensure identity at initialisation. The XLS-R backbone is frozen except for LoRA parameters, while the AASIST graph body remains trainable with a lower learning rate ($10^{-6}$); LoRA adapters and the classifier use $10^{-5}$. Training minimises a combined weighted BCE loss over all samples and a KL distillation term applied only to core samples against a frozen Stage-1 teacher ($T=2.0$, $\lambda=0.3$). Models are trained for 20 epochs with batch size 128 using Adam, balanced real/fake sampling, and gradient clipping at 1.0 on CrowdSource and US Congress with RawBoost augmentation. Both LoRA-only and merged checkpoints are saved for adapter-free inference, with further details provided in the appendix.

\section{Complete Experimental Results}
\subsection{Effect of Speakers on Detectability.}
Table~18 reports the average detection performance of baseline methods across multiple speaker variants of the same synthesis system (OpenVoice V2). We observe substantial performance variation across speakers for all methods, indicating that speaker characteristics alone can significantly affect deepfake detectability, even when the synthesis architecture is fixed. This suggests that system-level evaluation alone is insufficient, as intra-system speaker variability introduces additional domain shifts. While stronger models such as XLR-based detectors show more stable performance across speakers, no method is fully robust to speaker-induced variation. These results highlight the importance of evaluating deepfake detectors under multi-speaker settings and motivate the inclusion of speaker diversity when assessing open-world generalisation.
\begin{table}[h]
\begin{adjustbox}{width=0.5\columnwidth} 
\begin{tabular}{cccccc} \toprule
\textbf{Model}     & \textbf{OV S1} & \textbf{OV S2} & \textbf{OV S2} & \textbf{OV S3} & \textbf{OV S4} \\ \midrule
\textbf{RawNet2}   & 0.632          & 0.235          & 0.572          & 0.211          & 0.225          \\
\textbf{AASIST}    & 0.601          & 0.349          & 0.534          & 0.346          & 0.255          \\
\textbf{RawGAT-ST} & 0.691          & 0.266          & 0.685          & 0.183          & 0.279          \\
\textbf{PC-Dart}   & 0.545          & 0.383          & 0.559          & 0.363          & 0.333          \\
\textbf{SAMO}      & 0.695          & 0.499          & 0.731          & 0.451          & 0.503          \\
\textbf{NVA}       & 0.669          & 0.655          & 0.659          & 0.610          & 0.625          \\
\textbf{Purdue M2} & 0.405          & 0.529          & 0.515          & 0.542          & 0.514          \\
\textbf{XLR+R+A}   & 0.632          & 0.567          & 0.617          & 0.578          & 0.511          \\
\textbf{XLR+XLS}   & 0.369          & 0.411          & 0.449          & 0.417          & 0.320    \\ \bottomrule
\end{tabular}
\end{adjustbox}
\caption{Average performance of baseline methods across all subsets of OpenVoice V2 versions.}
\label{tab:ov_speaker}
\end{table}

\subsection{Detailed Zero-shot Performance}
We present selected results of open-world detection performance of the baseline methods on the Crowdsource and UC Congress subsets for both the TTS and Vocoder collections due to space limitations. Here, we provide the detailed results for all subsets on both collections.

\begin{table*}[h]
\begin{adjustbox}{width=\textwidth}
\begin{tabular}{ccccccccccccccccc} 
\toprule
\textbf{model}       & \textbf{CosyVoice} & \textbf{Zonos} & \textbf{Spark-TTS} & \textbf{F5-TTS} & \textbf{Fish-Speech} & \textbf{O.V s1} & \textbf{O.V s2} & \textbf{O.V s3} & \textbf{O.V s4} & \textbf{O.V s5} & \textbf{ChatTTS} & \textbf{Bark} & \textbf{XTTS} & \textbf{YourTTS} & \textbf{VITS} & \textbf{Avg} \\ \midrule
\textbf{XLS+R+A} & 0.436              & 0.297          & 0.565              & 0.676           & 0.163                & 0.566             & 0.367             & 0.491             & 0.411             & 0.293             & 0.180            & 0.180         & 0.200            & 0.135            & 0.469         & 0.362        \\
\textbf{XLR+SLS}        & 0.464              & 0.279          & 0.545              & 0.672           & 0.150                & 0.640             & 0.383             & 0.557             & 0.423             & 0.253             & 0.133            & 0.171         & 0.118            & 0.101            & 0.475         & 0.358        \\
\textbf{RawNet2}     & 0.220              & 0.470          & 0.252              & 0.716           & 0.447                & 0.820             & 0.314             & 0.867             & 0.260             & 0.395             & 0.550            & 0.522         & 0.287            & 0.127            & 0.375         & 0.441        \\
\textbf{ASSIST}      & 0.380              & 0.520          & 0.382              & 0.885           & 0.434                & 0.774             & 0.427             & 0.776             & 0.465             & 0.366             & 0.615            & 0.487         & 0.484            & 0.261            & 0.501         & 0.517        \\
\textbf{RawGAT-ST}   & 0.152              & 0.508          & 0.293              & 0.879           & 0.451                & 0.755             & 0.388             & 0.754             & 0.339             & 0.306             & 0.574            & 0.489         & 0.218            & 0.246            & 0.489         & 0.456        \\
\textbf{PC-Dart}     & 0.845              & 0.696          & 0.477              & 0.829           & 0.709                & 0.804             & 0.788             & 0.718             & 0.702             & 0.743             & 0.831            & 0.730         & 0.581            & 0.674            & 0.610         & 0.716        \\
\textbf{SAMO}        & 0.690              & 0.664          & 0.532              & 0.523           & 0.541                & 0.671             & 0.655             & 0.680             & 0.708             & 0.697             & 0.575            & 0.513         & 0.742            & 0.348            & 0.800         & 0.623        \\
\textbf{NVA}   & 0.590              & 0.601          & 0.552              & 0.568           & 0.519                & 0.592             & 0.598             & 0.569             & 0.593             & 0.607             & 0.344            & 0.334         & 0.471            & 0.224            & 0.612         & 0.518        \\
\textbf{Purdue-M2}      & 0.526              & 0.353          & 0.436              & 0.564           & 0.435                & 0.178             & 0.492             & 0.435             & 0.519             & 0.429             & 0.507            & 0.828         & 0.342            & 0.697            & 0.420         & 0.477     \\ \bottomrule 
\end{tabular}
\end{adjustbox}
\caption{Performance of the baseline methods on the Celebrity subset from the TTS collection in EER.}
\label{tab:apd:itwtts}
\end{table*}

\FloatBarrier

\begin{table*}[h]
\begin{adjustbox}{width=\textwidth}
\begin{tabular}{cccccccccccc}
\toprule
\textbf{Model}       & \textbf{BigVGAN} & \textbf{Vocos} & \textbf{BigVSAN} & \textbf{UnivNet} & \textbf{HiFi-GAN} & \textbf{MelGAN} & \textbf{MB Mel} & \textbf{FB Mel} & \textbf{Style Mel} & \textbf{PW GAN} & \textbf{Avg} \\ \midrule
\textbf{XLS+R+A} & 0.3820           & 0.2341         & 0.3150           & 0.1910           & 0.1398            & 0.0541          & 0.0656          & 0.0703          & 0.0675             & 0.066           & 0.159        \\
\textbf{XLR+SLS}        & 0.3827           & 0.2303         & 0.3147           & 0.1661           & 0.1196            & 0.0347          & 0.0476          & 0.0516          & 0.0520             & 0.0491          & 0.145        \\
\textbf{RawNet2}     & 0.4884           & 0.4861         & 0.5022           & 0.4565           & 0.4299            & 0.5224          & 0.4819          & 0.4124          & 0.3846             & 0.4769          & 0.464        \\
\textbf{ASSIST}      & 0.5089           & 0.5064         & 0.5362           & 0.5667           & 0.4773            & 0.4564          & 0.4515          & 0.4062          & 0.3811             & 0.5013          & 0.479        \\
\textbf{RawGAT-ST}   & 0.5083           & 0.4982         & 0.5256           & 0.5653           & 0.4821            & 0.5244          & 0.4990          & 0.4596          & 0.3909             & 0.5417          & 0.500        \\
\textbf{PC-Dart}     & 0.7381           & 0.7805         & 0.8424           & 0.7005           & 0.6718            & 0.6968          & 0.6407          & 0.7457          & 0.6129             & 0.6756          & 0.711        \\
\textbf{SAMO}        & 0.5081           & 0.5012         & 0.5302           & 0.6079           & 0.5262            & 0.5193          & 0.5350          & 0.4728          & 0.4531             & 0.5497          & 0.520        \\
\textbf{NVA}   & 0.5046           & 0.4929         & 0.5002           & 0.5254           & 0.5230            & 0.5004          & 0.4639          & 0.4710          & 0.5141             & 0.5292          & 0.502        \\
\textbf{Purdue-M2}      & 0.5101           & 0.5925         & 0.5303           & 0.5018           & 0.5851            & 0.7995          & 0.8013          & 0.7372          & 0.7481             & 0.7258          & 0.653      \\ \bottomrule
\end{tabular}
\end{adjustbox}
\caption{Performance of the baseline methods on the Celebrity subset from the Vocoder collection in EER.}
\label{tab:apd:itwvoc}
\end{table*}

\begin{table*}[h]
\begin{adjustbox}{width=\textwidth}
\begin{tabular}{ccccccccccccccccc}
\toprule
\textbf{model}       & \textbf{CosyVoice} & \textbf{Zonos} & \textbf{Spark-TTS} & \textbf{F5-TTS} & \textbf{Fish-Speech} & \textbf{O.V s1} & \textbf{O.V s2} & \textbf{O.V s3} & \textbf{O.V s4} & \textbf{O.V s5} & \textbf{ChatTTS} & \textbf{Bark} & \textbf{XTTS} & \textbf{YourTTS} & \textbf{VITS} & \textbf{Avg} \\ \midrule
\textbf{XLS+R+A} & 0.343                                      & 0.207          & 0.786              & 0.706                                   & 0.078                & 0.499             & 0.225             & 0.402             & 0.244             & 0.142             & 0.083            & 0.096         & 0.084            & 0.041            & 0.381         & 0.288        \\
\textbf{XLR+SLS}        & 0.418                                      & 0.261          & 0.642              & 0.600                                   & 0.134                & 0.524             & 0.337             & 0.455             & 0.336             & 0.202             & 0.101            & 0.135         & 0.083            & 0.061            & 0.396         & 0.312        \\
\textbf{RawNet2}     & 0.188                                      & 0.417          & 0.600              & 0.705                                   & 0.383                & 0.756             & 0.271             & 0.808             & 0.172             & 0.316             & 0.491            & 0.424         & 0.220            & 0.097            & 0.288         & 0.409        \\
\textbf{ASSIST}      & 0.451                                      & 0.531          & 0.661              & 0.928                                   & 0.484                & 0.721             & 0.479             & 0.738             & 0.474             & 0.429             & 0.633            & 0.462         & 0.503            & 0.358            & 0.490         & 0.556        \\
\textbf{RawGAT-ST}   & 0.157                                      & 0.509          & 0.634              & 0.898                                   & 0.456                & 0.684             & 0.394             & 0.728             & 0.279             & 0.294             & 0.548            & 0.424         & 0.198            & 0.232            & 0.431         & 0.458        \\
\textbf{PC-Dart}     & 0.634                                      & 0.432          & 0.306              & 0.600                                   & 0.451                & 0.546             & 0.501             & 0.597             & 0.409             & 0.424             & 0.640            & 0.481         & 0.442            & 0.648            & 0.451         & 0.504        \\
\textbf{SAMO}        & 0.777                                      & 0.728          & 0.629              & 0.579                                   & 0.624                & 0.771             & 0.739             & 0.770             & 0.763             & 0.781             & 0.661            & 0.514         & 0.819            & 0.410            & 0.867         & 0.695        \\
\textbf{NVA}   & 0.708                                      & 0.715          & 0.702              & 0.690                                   & 0.670                & 0.714             & 0.719             & 0.710             & 0.718             & 0.721             & 0.450            & 0.433         & 0.635            & 0.377            & 0.713         & 0.645        \\
\textbf{Purdue-M2}      & 0.316                                      & 0.175          & 0.163              & 0.289                                   & 0.228                & 0.064             & 0.257             & 0.188             & 0.254             & 0.229             & 0.274            & 0.687         & 0.170            & 0.481            & 0.199         & 0.265   \\ \bottomrule    
\end{tabular}
\end{adjustbox}
\caption{Performance of the baseline methods on the Crowdsource subset from the TTS collection in EER.}
\label{tab:apd:cvtts}
\end{table*}

\begin{table*}[h]
\begin{adjustbox}{width=\textwidth}
\begin{tabular}{cccccccccccc} \toprule
\textbf{Model}       & \textbf{BigVGAN} & \textbf{Vocos} & \textbf{BigVSAN} & \textbf{UnivNet} & \textbf{HiFi-GAN} & \textbf{MelGAN} & \textbf{MB Mel} & \textbf{FB Mel} & \textbf{Style Mel} & \textbf{PW GAN} & \textbf{Avg} \\ \midrule
\textbf{XLS+R+A} & 0.4100           & 0.2722         & 0.3410           & 0.1935              & 0.1077           & 0.0993          & 0.1224              & 0.1132              & 0.0783             & 0.1320       & 0.187        \\
\textbf{XLR+SLS}        & 0.4891           & 0.5169         & 0.3410           & 0.1752              & 0.4915           & 0.4643          & 0.5310              & 0.5081              & 0.3636             & 0.4389       & 0.432        \\
\textbf{RawNet2}     & 0.5053           & 0.4956         & 0.5136           & 0.5398              & 0.4989           & 0.4415          & 0.4896              & 0.5139              & 0.5241             & 0.4875       & 0.501        \\
\textbf{ASSIST}      & 0.4929           & 0.5129         & 0.4842           & 0.393               & 0.465            & 0.4862          & 0.4965              & 0.5464              & 0.5222             & 0.4365       & 0.484        \\
\textbf{RawGAT-ST}   & 0.4954           & 0.5314         & 0.4871           & 0.3842              & 0.4647           & 0.4481          & 0.4655              & 0.5210              & 0.4827             & 0.4083       & 0.469        \\
\textbf{PC-Dart}     & 0.5002           & 0.4940         & 0.5383           & 0.4697              & 0.5235           & 0.4429          & 0.4263              & 0.4631              & 0.4150             & 0.4292       & 0.470        \\
\textbf{SAMO}        & 0.5072           & 0.4901         & 0.5124           & 0.6392              & 0.5836           & 0.5808          & 0.611               & 0.5127              & 0.5150             & 0.6512       & 0.560        \\
\textbf{NVA}   & 0.4867           & 0.4675         & 0.4862           & 0.5747              & 0.6152           & 0.5873          & 0.5561              & 0.5972              & 0.6593             & 0.6149       & 0.565        \\
\textbf{Purdue-M2}      & 0.4847           & 0.5481         & 0.3067           & 0.3296              & 0.5649           & 0.7629          & 0.7532              & 0.7278              & 0.7001             & 0.7107       & 0.589   \\ \bottomrule
\end{tabular}
\end{adjustbox}
\caption{Performance of the baseline methods on the Crowdsource subset from the Vocoder collection in EER.}
\label{tab:apd:vcvoc}
\end{table*}

\begin{table*}[h]
\begin{adjustbox}{width=\textwidth}
\begin{tabular}{ccccccccccccccccl}\toprule
\textbf{model}       & \textbf{CosyVoice} & \textbf{Zonos} & \textbf{Spark-TTS} & \textbf{F5-TTS} & \textbf{Fish-Speech} & \multicolumn{1}{l}{\textbf{O.V s1}} & \multicolumn{1}{l}{\textbf{O.V s2}} & \multicolumn{1}{l}{\textbf{O.V s3}} & \multicolumn{1}{l}{\textbf{O.V s4}} & \multicolumn{1}{l}{\textbf{O.V s5}} & \textbf{ChatTTS} & \textbf{Bark} & \textbf{XTTS} & \textbf{YourTTS} & \textbf{VITS} & \multicolumn{1}{c}{\textbf{Avg}} \\ \midrule
\textbf{XLS+R+A} & 0.3253             & 0.554          & 0.833              & 0.871           & 0.339                & 0.748                                 & 0.607                                 & 0.723                                 & 0.651                                 & 0.469                                 & 0.380            & 0.296         & 0.349            & 0.255            & 0.753         & 0.544                            \\
\textbf{XLR+SLS}        & 0.7288             & 0.536          & 0.769              & 0.779           & 0.377                & 0.758                                 & 0.611                                 & 0.732                                 & 0.616                                 & 0.451                                 & 0.363            & 0.302         & 0.246            & 0.214            & 0.694         & 0.545                            \\
\textbf{RawNet2}     & 0.2093             & 0.440          & 0.205              & 0.686           & 0.412                & 0.775                                 & 0.264                                 & 0.761                                 & 0.195                                 & 0.326                                 & 0.521            & 0.461         & 0.232            & 0.103            & 0.284         & 0.392                            \\
\textbf{ASSIST}      & 0.5022             & 0.626          & 0.448              & 0.945           & 0.538                & 0.840                                 & 0.497                                 & 0.771                                 & 0.504                                 & 0.435                                 & 0.709            & 0.528         & 0.573            & 0.298            & 0.524         & 0.582                            \\
\textbf{RawGAT-ST}   & Chicken            & 0.591          & 0.303              & 0.936           & 0.568                & 0.817                                 & 0.413                                 & 0.703                                 & 0.323                                 & 0.357                                 & 0.692            & 0.557         & 0.224            & 0.270            & 0.515         & 0.519                            \\
\textbf{PC-Dart}     & 0.0483             & 0.053          & 0.018              & 0.037           & 0.069                & 0.050                                 & 0.026                                 & 0.051                                 & 0.017                                 & 0.022                                 & 0.237            & 0.091         & 0.031            & 0.085            & 0.028         & 0.058                            \\
\textbf{SAMO}        & 0.7851             & 0.703          & 0.599              & 0.567           & 0.610                & 0.741                                 & 0.704                                 & 0.755                                 & 0.707                                 & 0.737                                 & 0.657            & 0.566         & 0.789            & 0.382            & 0.833         & 0.676                            \\
\textbf{NVA}   & 0.7231             & 0.694          & 0.660              & 0.660           & 0.597                & 0.695                                 & 0.699                                 & 0.685                                 & 0.694                                 & 0.703                                 & 0.401            & 0.415         & 0.616            & 0.378            & 0.705         & 0.622                            \\
\textbf{Purdue-M2}      & 0.4867             & 0.274          & 0.440              & 0.545           & 0.363                & 0.108                                 & 0.477                                 & 0.409                                 & 0.497                                 & 0.398                                 & 0.478            & 0.812         & 0.305            & 0.699            & 0.381         & 0.445 \\ \bottomrule
\end{tabular}
\end{adjustbox}
\caption{Performance of the baselines methods on US Congress Subset inside the TTS partition.}
\end{table*}

\begin{table*}[h]
\begin{adjustbox}{width=\textwidth}
\begin{tabular}{cccccccccccc} \toprule
\textbf{Model}       & \textbf{BigVGAN}     & \textbf{Vocos}       & \textbf{BigVSAN}     & \textbf{UnivNet}     & \textbf{HiFi-GAN}    & \textbf{MelGAN}      & \textbf{MB Mel}      & \textbf{FB Mel}      & \textbf{Style Mel}   & \textbf{PW GAN}      & \textbf{Avg}         \\ \midrule
\textbf{XLS+R+A} & 0.428                & 0.357                & 0.396                & 0.250                & 0.223                & 0.171                & 0.208                & 0.197                & 0.202                & 0.191                & 0.262                \\
\textbf{XLR+SLS}        & 0.417                & 0.318                & 0.380                & 0.180                & 0.152                & 0.098                & 0.127                & 0.102                & 0.104                & 0.108                & 0.199                \\
\textbf{RawNet2}     & 0.464                & 0.461                & 0.468                & 0.352                & 0.351                & 0.430                & 0.408                & 0.341                & 0.276                & 0.396                & 0.395                \\
\textbf{ASSIST}      & 0.497                & 0.501                & 0.494                & 0.606                & 0.643                & 0.649                & 0.637                & 0.689                & 0.709                & 0.582                & 0.600                \\
\textbf{RawGAT-ST}   & 0.506                & 0.492                & 0.512                & 0.473                & 0.428                & 0.443                & 0.433                & 0.425                & 0.376                & 0.492                & 0.458                \\
\textbf{PC-Dart}     & 0.028                & 0.032                & 0.036                & 0.024                & 0.026                & 0.024                & 0.023                & 0.028                & 0.021                & 0.022                & 0.026                \\
\textbf{SAMO}        & 0.505                & 0.493                & 0.511                & 0.462                & 0.445                & 0.500                & 0.489                & 0.418                & 0.453                & 0.502                & 0.478                \\
\textbf{NVA}   & 0.493                & 0.492                & 0.487                & 0.407                & 0.408                & 0.382                & 0.354                & 0.364                & 0.401                & 0.433                & 0.422                \\
\textbf{Purdue-M2}      & 0.500                & 0.538                & 0.504                & 0.539                & 0.650                & 0.812                & 0.765                & 0.745                & 0.792                & 0.769                & 0.661                \\ \bottomrule
\end{tabular}
\end{adjustbox}
\caption{Performance of the baseline methods on the US Congress subset from the Vocoder collection in EER.}
\end{table*}

\begin{table*}[h]
\begin{adjustbox}{width=\textwidth}
\begin{tabular}{ccccccccccccccccc}
\toprule
\textbf{Model}       & \textbf{CosyVoice} & \textbf{Zonos} & \textbf{Spark-TTS} & \textbf{F5-TTS} & \textbf{Fish-Speech} & \textbf{O.V s1} & \textbf{O.V s2} & \textbf{O.V s3} & \textbf{O.V s4} & \textbf{O.V s5} & \textbf{ChatTTS} & \textbf{Bark} & \textbf{XTTS} & \textbf{YourTTS} & \textbf{VITS} & \textbf{Avg} \\ \midrule
\textbf{XLS+R+A} & 0.542              & 0.382          & 0.600              & 0.630           & 0.268                & 0.472             & 0.344             & 0.467             & 0.348             & 0.247             & 0.220            & 0.166         & 0.198            & 0.138            & 0.459         & 0.365        \\
\textbf{XLR+SLS}        & 0.542              & 0.263          & 0.514              & 0.598           & 0.208                & 0.474             & 0.284             & 0.441             & 0.280             & 0.171             & 0.126            & 0.103         & 0.080            & 0.044            & 0.363         & 0.299        \\
\textbf{RawNet2}     & 0.069              & 0.244          & 0.043              & 0.627           & 0.420                & 0.708             & 0.079             & 0.588             & 0.017             & 0.151             & 0.449            & 0.320         & 0.056            & 0.008            & 0.076         & 0.257        \\
\textbf{ASSIST}      & 0.262              & 0.324          & 0.163              & 0.934           & 0.467                & 0.517             & 0.198             & 0.385             & 0.149             & 0.138             & 0.534            & 0.276         & 0.303            & 0.078            & 0.156         & 0.325        \\
\textbf{RawGAT-ST}   & 0.087              & 0.307          & 0.094              & 0.875           & 0.597                & 0.503             & 0.086             & 0.336             & 0.019             & 0.112             & 0.467            & 0.315         & 0.041            & 0.024            & 0.133         & 0.266        \\
\textbf{PC-Dart}     & 0.194              & 0.125          & 0.081              & 0.152           & 0.237                & 0.140             & 0.125             & 0.126             & 0.105             & 0.112             & 0.390            & 0.176         & 0.154            & 0.270            & 0.149         & 0.169        \\
\textbf{SAMO}        & 0.682              & 0.527          & 0.349              & 0.206           & 0.465                & 0.571             & 0.473             & 0.674             & 0.424             & 0.630             & 0.477            & 0.346         & 0.720            & 0.120            & 0.724         & 0.492        \\
\textbf{NVA}   & 0.604              & 0.595          & 0.572              & 0.566           & 0.504                & 0.611             & 0.611             & 0.610             & 0.610             & 0.611             & 0.289            & 0.365         & 0.554            & 0.281            & 0.611         & 0.533        \\
\textbf{Purdue-M2}      & 0.440              & 0.322          & 0.412              & 0.484           & 0.424                & 0.114             & 0.464             & 0.420             & 0.445             & 0.349             & 0.426            & 0.829         & 0.343            & 0.723            & 0.381         & 0.438 \\ \bottomrule      
\end{tabular}
\end{adjustbox}
\caption{Performance of the baselines methods on Audiobook Subset inside the TTS partition.}
\end{table*}

\begin{table*}[h]
\begin{adjustbox}{width=\textwidth}
\begin{tabular}{cccccccccccc}
\toprule
\textbf{Model}       & \textbf{BigVGAN}     & \textbf{Vocos}       & \textbf{BigVSAN}     & \textbf{UnivNet}     & \textbf{HiFi-GAN}    & \textbf{MelGAN}      & \textbf{MB Mel}      & \textbf{FB Mel}      & \textbf{Style Mel}   & \textbf{PW GAN}      & \textbf{Avg}         \\ \midrule
\textbf{XLS+R+A} & 0.389                & 0.243                & 0.344                & 0.198                & 0.156                & 0.140                & 0.127                & 0.116                & 0.088                & 0.126                & 0.192                \\
\textbf{XLR+SLS}        & 0.382                & 0.208                & 0.338                & 0.159                & 0.122                & 0.043                & 0.050                & 0.061                & 0.052                & 0.061                & 0.148                \\
\textbf{RawNet2}     & 0.475                & 0.424                & 0.488                & 0.436                & 0.438                & 0.461                & 0.427                & 0.369                & 0.373                & 0.455                & 0.434                \\
\textbf{ASSIST}      & 0.503                & 0.474                & 0.506                & 0.531                & 0.489                & 0.404                & 0.404                & 0.388                & 0.405                & 0.432                & 0.454                \\
\textbf{RawGAT-ST}   & 0.517                & 0.485                & 0.517                & 0.498                & 0.494                & 0.493                & 0.446                & 0.412                & 0.418                & 0.473                & 0.475                \\
\textbf{PC-Dart}     & 0.217                & 0.238                & 0.253                & 0.195                & 0.202                & 0.163                & 0.157                & 0.177                & 0.164                & 0.151                & 0.192                \\
\textbf{SAMO}        & 0.489                & 0.495                & 0.493                & 0.501                & 0.503                & 0.510                & 0.538                & 0.500                & 0.477                & 0.553                & 0.506                \\
\textbf{NVA}   & 0.506                & 0.512                & 0.521                & 0.542                & 0.574                & 0.562                & 0.548                & 0.570                & 0.590                & 0.584                & 0.551                \\
\textbf{Purdue-M2}      & 0.462                & 0.507                & 0.476                & 0.470                & 0.477                & 0.697                & 0.681                & 0.627                & 0.631                & 0.612                & 0.564                \\ \bottomrule
\end{tabular}
\end{adjustbox}
\caption{Performance of the baseline methods on the Audiobook subset from the Vocoder collection in EER.}
\label{tab:apd:mlsvoc}
\end{table*}
\FloatBarrier

\vspace{2cm}

\subsection{Single System Generalisation Result}
In this section, we present the full single system generalisation results.
\subsubsection{Same textual information, same domain.}
This section provides complete results for Section \ref{sec:single}.
\begin{table}[h]
\begin{adjustbox}{width=\textwidth}
\begin{tabular}{ccccccccccccccccc} \toprule
Model        & CosyVoice & Zonos & Spark-TTS & F5-TTS & Fish-Speech & \multicolumn{1}{l}{O.V s1} & \multicolumn{1}{l}{O.V s2} & \multicolumn{1}{l}{O.V s3} & \multicolumn{1}{l}{O.V s4} & \multicolumn{1}{l}{O.V s5} & ChatTTS & Bark  & XTTS & YourTTS & VITS  & Avg   \\ \midrule
Spark-TTS     & 0.006     & 0.000 & 0.000     & 0.024  & 0.116       & 0.314                        & 0.027                        & 0.400                        & 0.000                        & 0.004                        & 0.524   & 0.019 & 0.000   & 0.000   & 0.001 & 0.096 \\
F5\_tts      & 0.219     & 0.003 & 0.082     & 0.000  & 0.209       & 0.288                        & 0.116                        & 0.444                        & 0.032                        & 0.290                        & 0.442   & 0.028 & 0.003   & 0.008   & 0.079 & 0.150 \\
fish\_speech & 0.005     & 0.001 & 0.171     & 0.427  & 0.000       & 0.252                        & 0.001                        & 0.190                        & 0.000                        & 0.000                        & 0.351   & 0.002 & 0.000   & 0.000   & 0.000 & 0.093 \\
XTTS         & 0.112     & 0.007 & 0.242     & 0.517  & 0.229       & 0.292                        & 0.035                        & 0.318                        & 0.068                        & 0.097                        & 0.581   & 0.055 & 0.000   & 0.008   & 0.024 & 0.172 \\
VITS         & 0.004     & 0.000 & 0.071     & 0.349  & 0.090       & 0.244                        & 0.002                        & 0.279                        & 0.000                        & 0.001                        & 0.574   & 0.012 & 0.000   & 0.000   & 0.000 & 0.108 \\ \bottomrule
\end{tabular}
\end{adjustbox}
\caption{Single system generalisation performance for models trained on real audios from the Crowdsource train partition and their corresponding fake audios generated by a single TTS model in ERR, tested on the Crowdsource train partition of the TTS collection.}
\label{tab:apd24}
\end{table}

\begin{table}[h]
\begin{tabular}{cccccccccccc} \toprule
Train   Patt. & BigVGAN & Vocos & BigVSAN & UnivNet & HiFi-GAN & MelGAN & MB Mel & FB Mel & Style Mel & PW GAN & Avg   \\ \midrule
Spark-TTS     & 0.501   & 0.502 & 0.499   & 0.497   & 0.479    & 0.498  & 0.502  & 0.497  & 0.477     & 0.509  & 0.496 \\
F5-TTS        & 0.498   & 0.496 & 0.492   & 0.499   & 0.470    & 0.498  & 0.496  & 0.495  & 0.463     & 0.513  & 0.492 \\
Fish-speech   & 0.493   & 0.487 & 0.489   & 0.462   & 0.359    & 0.410  & 0.435  & 0.431  & 0.369     & 0.449  & 0.438 \\
XTTS          & 0.499   & 0.499 & 0.498   & 0.506   & 0.510    & 0.487  & 0.491  & 0.488  & 0.484     & 0.496  & 0.496 \\
VITS          & 0.500   & 0.500 & 0.498   & 0.501   & 0.446    & 0.483  & 0.486  & 0.483  & 0.454     & 0.500  & 0.485 \\ \bottomrule
\end{tabular}
\caption{Single system generalisation performance for models trained on real audios from the Crowdsource train partition and their corresponding fake audios generated by a single TTS model in ERR, tested on the Crowdsource train partition of the vocoder collection.}
\end{table}
\FloatBarrier

\begin{table}[h]
\begin{adjustbox}{width=\textwidth}
\begin{tabular}{ccccccccccccccccc} \toprule
Model        & CosyVoice & Zonos & Spark-TTS & F5-TTS & Fish-Speech & \multicolumn{1}{l}{O.V s1} & \multicolumn{1}{l}{O.V s2} & \multicolumn{1}{l}{O.V s3} & \multicolumn{1}{l}{O.V s4} & \multicolumn{1}{l}{O.V s5} & ChatTTS & Bark  & XTTS & YourTTS & VITS  & Avg   \\ \midrule
Mel GAN     & 0.678     & 0.606 & 0.675     & 0.435  & 0.465       & 0.411                        & 0.588                        & 0.541                        & 0.666                        & 0.629                        & 0.148   & 0.559 & 0.640   & 0.647   & 0.522 & 0.547 \\
HiFi-GAN     & 0.261     & 0.001 & 0.232     & 0.582  & 0.205       & 0.428                        & 0.316                        & 0.446                        & 0.221                        & 0.324                        & 0.527   & 0.003 & 0.001   & 0.000   & 0.025 & 0.238 \\
Vocos & 0.277     & 0.274 & 0.386     & 0.104  & 0.106       & 0.305                        & 0.358                        & 0.494                        & 0.478                        & 0.393                        & 0.050   & 0.106 & 0.209   & 0.062   & 0.321 & 0.262 \\
UnivNet         & 0.186     & 0.001 & 0.232     & 0.582  & 0.205       & 0.428                        & 0.316                        & 0.446                        & 0.221                        & 0.324                        & 0.527   & 0.003 & 0.001   & 0.000   & 0.025 & 0.233 \\
Mel GAN         & 0.221     & 0.134 & 0.282     & 0.441  & 0.200       & 0.359                        & 0.188                        & 0.433                        & 0.170                        & 0.281                        & 0.302   & 0.150 & 0.095   & 0.107   & 0.158 & 0.235 \\ \bottomrule
\end{tabular}
\end{adjustbox}
\caption{Single system generalisation performance for models trained on real audios from the Crowdsource train partition and their corresponding fake audios generated by a single vocoder model in ERR, tested on the Crowdsource train partition of the TTS collection.}
\end{table}

\begin{table}[h]
\begin{adjustbox}{width=\textwidth}
\begin{tabular}{cccccccccccc} \toprule
Train   Patt. & BigVGAN & Vocos & BigVSAN & UnivNet & HiFi-GAN & MelGAN & MB Mel & FB Mel & Style Mel & PW GAN & Avg   \\  \midrule
Spark-TTS     & 0.250   & 0.241 & 0.183   & 0.130   & 0.400    & 0.262  & 0.276  & 0.287  & 0.341     & 0.328  & 0.270 \\
F5-TTS        & 0.490   & 0.476 & 0.479   & 0.433   & 0.000    & 0.277  & 0.232  & 0.195  & 0.019     & 0.245  & 0.285 \\
Fish-speech   & 0.273   & 0.016 & 0.119   & 0.024   & 0.063    & 0.018  & 0.021  & 0.024  & 0.031     & 0.031  & 0.062 \\
XTTS          & 0.490   & 0.476 & 0.479   & 0.433   & 0.000    & 0.277  & 0.232  & 0.195  & 0.019     & 0.245  & 0.285 \\
VITS          & 0.489   & 0.467 & 0.477   & 0.351   & 0.163    & 0.000  & 0.006  & 0.015  & 0.012     & 0.012  & 0.199 \\\bottomrule 
\end{tabular}
\end{adjustbox}
\caption{Single system generalisation performance for models trained on real audios from the Crowdsource train partition and their corresponding fake audios generated by a single vocoder model in ERR, tested on the Crowdsource train partition of the vocoder collection.}
\end{table}
\FloatBarrier

\subsubsection{Same domain different textual information.} This section provides complete results for Section \ref{sec:single}.

\begin{table}[h]
\begin{adjustbox}{width=\textwidth}
\begin{tabular}{ccccccccccccccccc} \toprule
Model        & CosyVoice & Zonos & Spark-TTS & F5-TTS & Fish-Speech & \multicolumn{1}{l}{O.V s1} & \multicolumn{1}{l}{O.V s2} & \multicolumn{1}{l}{O.V s3} & \multicolumn{1}{l}{O.V s4} & \multicolumn{1}{l}{O.V s5} & ChatTTS & Bark  & XTTS & YourTTS & VITS  & Avg   \\ \midrule
Spark-TTS     & 0.007     & 0.001 & 0.001     & 0.030  & 0.126       & 0.329                        & 0.032                        & 0.407                        & 0.001                        & 0.005                        & 0.528   & 0.021 & 0.001   & 0.001   & 0.005 & 0.100 \\
F5\_tts      & 0.229     & 0.004 & 0.091     & 0.000  & 0.219       & 0.300                        & 0.128                        & 0.453                        & 0.035                        & 0.296                        & 0.446   & 0.029 & 0.004   & 0.008   & 0.080 & 0.155 \\
Fish-Speech  & 0.006     & 0.001 & 0.176     & 0.437  & 0.001       & 0.268                        & 0.001                        & 0.195                        & 0.000                        & 0.001                        & 0.363   & 0.003 & 0.000   & 0.000   & 0.006 & 0.097 \\
XTTS         & 0.118     & 0.008 & 0.246     & 0.519  & 0.240       & 0.304                        & 0.040                        & 0.327                        & 0.077                        & 0.105                        & 0.586   & 0.061 & 0.000   & 0.007   & 0.027 & 0.178 \\
VITS         & 0.005     & 0.001 & 0.076     & 0.351  & 0.101       & 0.255                        & 0.003                        & 0.282                        & 0.001                        & 0.002                        & 0.572   & 0.015 & 0.000   & 0.000   & 0.001 & 0.111 \\ \bottomrule 
\end{tabular} 
\end{adjustbox}
\caption{Single system generalisation performance for models trained on real audios from the Crowdsource train partition and their corresponding fake audios generated by a single TTS model in ERR, tested on the Crowdsource test partition of the TTS collection.}
\end{table}

\begin{table}[h]
\begin{tabular}{cccccccccccc} \toprule
Model & BigVGAN & Vocos & BigVSAN & UnivNet & HiFi-GAN & MelGAN & MB Mel & FB Mel & Style Mel & PW GAN & Avg   \\ \midrule
Spark-TTS     & 0.500   & 0.503 & 0.499   & 0.497   & 0.483    & 0.496  & 0.503  & 0.495  & 0.473     & 0.511  & 0.496 \\
F5-TTS        & 0.496   & 0.495 & 0.491   & 0.498   & 0.474    & 0.500  & 0.496  & 0.495  & 0.457     & 0.513  & 0.491 \\
Fish-Speech   & 0.491   & 0.489 & 0.489   & 0.466   & 0.366    & 0.419  & 0.440  & 0.440  & 0.365     & 0.455  & 0.442 \\
XTTS          & 0.499   & 0.501 & 0.498   & 0.507   & 0.508    & 0.489  & 0.491  & 0.486  & 0.481     & 0.500  & 0.496 \\
VITS          & 0.500   & 0.502 & 0.500   & 0.500   & 0.449    & 0.484  & 0.489  & 0.484  & 0.447     & 0.501  & 0.485 \\ \bottomrule
\end{tabular}
\caption{Single system generalisation performance for models trained on real audios from the Crowdsource train partition and their corresponding fake audios generated by a single TTS model in ERR, tested on the Crowdsource test partition of the vocoder collection.}

\end{table}

\begin{table}[h]
\begin{adjustbox}{width=\textwidth}
\begin{tabular}{ccccccccccccccccc} \toprule
Model        & CosyVoice & Zonos & Spark-TTS & F5-TTS & Fish-Speech & \multicolumn{1}{l}{O.V s1} & \multicolumn{1}{l}{O.V s2} & \multicolumn{1}{l}{O.V s3} & \multicolumn{1}{l}{O.V s4} & \multicolumn{1}{l}{O.V s5} & ChatTTS & Bark  & XTTS & YourTTS & VITS  & Avg   \\ \midrule
BigVGAN     & 0.695     & 0.625 & 0.687     & 0.450  & 0.489       & 0.433                        & 0.613                        & 0.569                        & 0.677                        & 0.635                        & 0.157   & 0.579 & 0.662   & 0.669   & 0.531 & 0.565 \\
HiFi-GAN    & 0.277     & 0.002 & 0.244     & 0.587  & 0.219       & 0.453                        & 0.340                        & 0.464                        & 0.234                        & 0.345                        & 0.544   & 0.004 & 0.001   & 0.000   & 0.029 & 0.249 \\
Vocos & 0.320     & 0.325 & 0.428     & 0.140  & 0.148       & 0.336                        & 0.402                        & 0.522                        & 0.514                        & 0.433                        & 0.084   & 0.146 & 0.261   & 0.099   & 0.370 & 0.302 \\
UnivNet         & 0.215     & 0.144 & 0.421     & 0.127  & 0.093       & 0.437                        & 0.237                        & 0.281                        & 0.170                        & 0.299                        & 0.044   & 0.182 & 0.085   & 0.105   & 0.336 & 0.212 \\
Mel GAN         & 0.227     & 0.147 & 0.287     & 0.447  & 0.213       & 0.371                        & 0.202                        & 0.439                        & 0.181                        & 0.284                        & 0.316   & 0.163 & 0.112   & 0.120   & 0.164 & 0.245 \\ \bottomrule
\end{tabular}
\end{adjustbox}
\caption{Single system generalisation performance for models trained on real audios from the Crowdsource train partition and their corresponding fake audios generated by a single vocoder in ERR, tested on the Crowdsource test partition of the TTS collection.}
\end{table}

\begin{table}[h]
\begin{tabular}{cccccccccccc} \toprule
Train   Patt. & BigVGAN & Vocos & BigVSAN & UnivNet & HiFi-GAN & MelGAN & MB Mel & FB Mel & Style Mel & PW GAN & Avg   \\ \midrule  
BigVGAN    & 0.281   & 0.274 & 0.211   & 0.148   & 0.419    & 0.292  & 0.304  & 0.314  & 0.370     & 0.350  & 0.296 \\
HiFi-GAN         & 0.491   & 0.479 & 0.483   & 0.438   & 0.001    & 0.292  & 0.239  & 0.203  & 0.019     & 0.252  & 0.290 \\
Vocos   & 0.287   & 0.049 & 0.153   & 0.050   & 0.099    & 0.041  & 0.045  & 0.049  & 0.063     & 0.065  & 0.090 \\
UnivNet           & 0.435   & 0.324 & 0.260   & 0.012   & 0.101    & 0.012  & 0.032  & 0.033  & 0.025     & 0.028  & 0.126 \\
Mel GAN         & 0.486   & 0.466 & 0.474   & 0.345   & 0.176    & 0.002  & 0.010  & 0.020  & 0.016     & 0.016  & 0.201 \\ \bottomrule
\end{tabular} 
\caption{Single system generalisation performance for models trained on real audios from the Crowdsource train partition and their corresponding fake audios generated by a single vocoder in ERR, tested on the Crowdsource test partition of the vocoder collection.}
\end{table}
\FloatBarrier

\subsubsection{Different domain and different textual information.}
This section provides complete results for Section \ref{sec:single}.

\begin{table}[h]
\begin{adjustbox}{width=\textwidth}
\begin{tabular}{ccccccccccccccccc} \toprule
Model        & CosyVoice & Zonos & Spark-TTS & F5-TTS & Fish-Speech & \multicolumn{1}{l}{O.V s1} & \multicolumn{1}{l}{O.V s2} & \multicolumn{1}{l}{O.V s3} & \multicolumn{1}{l}{O.V s4} & \multicolumn{1}{l}{O.V s5} & ChatTTS & Bark  & XTTS & YourTTS & VITS  & Avg   \\ \midrule
Spark-TTS     & 0.301     & 0.307 & 0.320     & 0.542  & 0.473       & 0.998                        & 0.379                        & 0.998                        & 0.246                        & 0.336                        & 0.979   & 0.507 & 0.332   & 0.291   & 0.217 & 0.482 \\
F5\_tts      & 0.844     & 0.341 & 0.715     & 0.004  & 0.611       & 0.987                        & 0.918                        & 0.991                        & 0.757                        & 0.981                        & 0.943   & 0.582 & 0.208   & 0.278   & 0.740 & 0.660 \\
fish\_speech & 0.400     & 0.359 & 0.459     & 0.948  & 0.372       & 0.980                        & 0.334                        & 0.878                        & 0.277                        & 0.291                        & 0.935   & 0.463 & 0.370   & 0.358   & 0.232 & 0.510 \\
XTTS         & 0.684     & 0.286 & 0.656     & 0.985  & 0.599       & 0.992                        & 0.895                        & 0.992                        & 0.670                        & 0.980                        & 0.993   & 0.533 & 0.053   & 0.205   & 0.519 & 0.669 \\
VITS         & 0.301     & 0.270 & 0.375     & 0.938  & 0.464       & 0.998                        & 0.188                        & 0.989                        & 0.129                        & 0.162                        & 0.991   & 0.503 & 0.202   & 0.246   & 0.067 & 0.455 \\ \bottomrule  
\end{tabular}
\end{adjustbox}
\caption{Single system generalisation performance for models trained on real audios from the Crowdsource train partition and their corresponding fake audios generated by a single TTS model in ERR, tested on the US Congress test partition of the TTS collection.}
\end{table}

\begin{table}[h]
\begin{adjustbox}{width=\textwidth}
\begin{tabular}{cccccccccccc}
\toprule
Model        & BigVGAN & Vocos  & BigVSAN & UnivNet & HiFi-GAN & MelGAN & MB Mel & FB Mel & Style Mel & PW GAN & Avg   \\\hline
sparktts     & 0.527   & 0.532  & 0.522   & 0.516   & 0.533    & 0.573  & 0.557  & 0.570  & 0.531     & 0.554  & 0.541 \\
f5\_tts      & 0.569   & 0.568  & 0.561   & 0.537   & 0.559    & 0.590  & 0.572  & 0.574  & 0.565     & 0.598  & 0.569 \\
fish\_speech & 0.474   & 0.469  & 0.470   & 0.458   & 0.477    & 0.440  & 0.473  & 0.493  & 0.440     & 0.447  & 0.464 \\
xtts         & 0.547   & 0.551  & 0.534   & 0.494   & 0.519    & 0.537  & 0.540  & 0.551  & 0.509     & 0.524  & 0.531 \\
vits         & 0.5263  & 0.5264 & 0.5179  & 0.5064  & 0.5388   & 0.5454 & 0.5489 & 0.5591 & 0.5188    & 0.5422 & 0.533\\ \bottomrule
\end{tabular}   
\end{adjustbox}
\caption{Single system generalisation performance for models trained on real audios from the Crowdsource train partition and their corresponding fake audios generated by a single TTS model in ERR, tested on the US Congress test partition of the vocoder collection.}
\end{table}

\begin{table}[h]
\begin{adjustbox}{width=\textwidth}
\begin{tabular}{ccccccccccccccccc} \toprule
Model        & CosyVoice & Zonos & Spark-TTS & F5-TTS & Fish-Speech & \multicolumn{1}{l}{O.V s1} & \multicolumn{1}{l}{O.V s2} & \multicolumn{1}{l}{O.V s3} & \multicolumn{1}{l}{O.V s4} & \multicolumn{1}{l}{O.V s5} & ChatTTS & Bark  & XTTS & YourTTS & VITS  & Avg   \\ \midrule
BigVGAN    & 0.608     & 0.537 & 0.673     & 0.351  & 0.398       & 0.325                        & 0.537                        & 0.523                        & 0.647                        & 0.622                        & 0.119   & 0.458 & 0.557   & 0.563   & 0.508 & 0.495 \\
F5\_tts      & 0.644     & 0.483 & 0.652     & 0.982  & 0.528       & 0.999                        & 0.900                        & 0.999                        & 0.550                        & 0.917                        & 0.989   & 0.522 & 0.432   & 0.381   & 0.387 & 0.691 \\
fish\_speech & 0.689     & 0.741 & 0.858     & 0.337  & 0.369       & 0.585                        & 0.857                        & 0.813                        & 0.874                        & 0.833                        & 0.239   & 0.380 & 0.531   & 0.256   & 0.832 & 0.613 \\
XTTS         & 0.458     & 0.288 & 0.704     & 0.283  & 0.210       & 0.711                        & 0.481                        & 0.479                        & 0.353                        & 0.551                        & 0.108   & 0.358 & 0.175   & 0.226   & 0.606 & 0.399 \\
VITS         & 0.726     & 0.669 & 0.751     & 0.855  & 0.579       & 0.826                        & 0.695                        & 0.844                        & 0.675                        & 0.775                        & 0.639   & 0.586 & 0.573   & 0.586   & 0.671 & 0.697 \\ \bottomrule 
\end{tabular}
\end{adjustbox}
\caption{Single system generalisation performance for models trained on real audios from the Crowdsource train partition and their corresponding fake audios generated by a single vocoder in ERR, tested on the US Congress test partition of the TTS collection.}
\end{table}

\begin{table}[]
\begin{tabular}{cccccccccccc}
\toprule
Model    & BigVGAN & Vocos & BigVSAN & UnivNet & HiFi-GAN & MelGAN & MB Mel & FB Mel & Style Mel & PW GAN & Avg   \\\hline
BigVGAN  & 0.440   & 0.401 & 0.377   & 0.343   & 0.409    & 0.426  & 0.382  & 0.383  & 0.395     & 0.411  & 0.426 \\
HiFi-GAN & 0.486   & 0.468 & 0.471   & 0.434   & 0.444    & 0.365  & 0.407  & 0.436  & 0.383     & 0.384  & 0.441 \\
Vocos    & 0.398   & 0.217 & 0.301   & 0.240   & 0.229    & 0.216  & 0.205  & 0.215  & 0.207     & 0.227  & 0.320 \\
UnivNet  & 0.427   & 0.346 & 0.272   & 0.077   & 0.146    & 0.070  & 0.093  & 0.104  & 0.105     & 0.087  & 0.227 \\
Mel GAN  & 0.485   & 0.423 & 0.456   & 0.290   & 0.149    & 0.037  & 0.065  & 0.083  & 0.084     & 0.065  & 0.315\\
\bottomrule
\end{tabular}
\caption{Single system generalisation performance for models trained on real audios from the Crowdsource train partition and their corresponding fake audios generated by a single vocoder in ERR, tested on the US Congress test partition of the vocoder collection.}
\label{tab:apd35}
\end{table}

\end{document}